\documentclass{article}

\usepackage[english]{babel}

\usepackage[letterpaper,top=2cm,bottom=2cm,left=3cm,right=3cm,marginparwidth=1.75cm]{geometry}

\usepackage{amsmath,amsfonts,amssymb}
\usepackage{graphicx}
\usepackage{algorithm}
\usepackage{algpseudocode}
\usepackage{multirow}
\usepackage{longtable}
\usepackage[colorlinks=true, allcolors=blue]{hyperref}
\allowdisplaybreaks

\newcommand{\R}{\mathbb{R}}
\newcommand{\Z}{\mathbb{Z}}
\renewcommand{\S}{\mathbb{S}}
\newcommand{\cX}{\mathcal{X}}
\newcommand{\cS}{\mathcal{S}}
\newcommand{\cT}{\mathcal{T}}
\newcommand{\cF}{\mathcal{F}}
\newcommand{\norm}[1]{\left\lVert #1 \right\rVert}
\newcommand{\inprod}[2]{\left\langle #1, #2\right\rangle}
\newtheorem{lemma}{Lemma}
\newtheorem{remark}{Remark}
\newtheorem{definition}{Definition}
\newtheorem{theorem}{Theorem}
\newtheorem{example}{Example}

\title{PNOD: An Efficient Projected Newton Framework for Exact Optimal Experimental Designs}
\author{Ling Liang \thanks{Department of Mathematics, University of Maryland, College Park. (\url{liang.ling@u.nus.edu})} \and Haizhao Yang \thanks{Department of Mathematics and Department of Computer Science, University of Maryland, College Park. (\url{hzyang@umd.edu})}}
\date{}

\begin{document}
\maketitle

\begin{abstract}
Computing the exact optimal experimental design has been a longstanding challenge in various scientific fields. This problem, when formulated using a specific information function, becomes a mixed-integer nonlinear programming (MINLP) problem, which is typically NP-hard, thus making the computation of a globally optimal solution extremely difficult. The branch and bound (BnB) method is a widely used approach for solving such MINLPs, but its practical efficiency heavily relies on the ability to solve continuous relaxations effectively within the BnB search tree. In this paper, we propose a novel projected Newton framework, combining with a vertex exchange method for efficiently solving the associated subproblems, designed to enhance the BnB method. This framework offers strong convergence guarantees by utilizing recent advances in solving self-concordant optimization and convex quadratic programming problems. Extensive numerical experiments on A-optimal and D-optimal design problems, two of the most commonly used models, demonstrate the framework's promising numerical performance. Specifically, our framework significantly improves the efficiency of node evaluation within the BnB search tree and enhances the accuracy of solutions compared to state-of-the-art methods. The proposed framework is implemented in an open source Julia package called \texttt{PNOD.jl}, which opens up possibilities for its application in a wide range of real-world scenarios.
\end{abstract}

\textbf{Keywords:} Exact optimal experimental design, Mixed-integer nonlinear programming, Branch and bound method, Projected Newton method, Vertex exchange method.

\section{Introduction}
The linear regression model of the following form:
\begin{equation}
    \label{eq:linearregression}
    \min_{\beta \in\mathbb{R}^n} \; \norm{A\beta  - y}_2,
\end{equation}
has been one of the most fundamental models in statistics \cite{montgomery2021introduction}, where $\beta\in\R^n$ stands for the parameters to be estimated, $A:= [a_1^T; \dots; a_m^T]\in \R^{m\times n}$ consists of $m$ rows, each of which represents an individual experiment, and $y\in \R^m$ captures the outcome of the experiments. Throughout this paper, we assume that the matrix $A$ has full column rank. In practical applications, $n$ is much smaller than $m$, i.e., the size of the parameter is significantly smaller than the total number of experiments that one can conduct in principle. However, it is typically the case that conducting a single experiment can be costly, making the whole process of solving the linear regression model \eqref{eq:linearregression} extremely time-consuming. 

To address the aforementioned issue, a commonly used approach is to first solve the associated optimal experimental design problem (OEDP), which essentially seeks a subset of $N$ (here $N$ is prespecified, and we require $N \geq n$ for ensuring the solvability of the linear model \eqref{eq:linearregression} and $N \ll m$ to save computational costs) experiments with the most information over the entire experiment space \cite{pukelsheim2006optimal, fedorov2013theory}. The selected subset of experiments can thus be represented by a vector of nonnegative integers $x:=(x_1;\dots;x_m)\in\mathbb{R}^m$, where $x_i$ denotes the number of repetitions for the $i$-th experiment $a_i$, for $i = 1,\dots, m$. To quantify the information mathematically, one relies on a certain information function $\phi:\S_{+}^n\to \R $ with respect to the information matrix associated with $A$ and $x$ that is defined as $X(x):= \sum_{i = 1}^m x_ia_ia_i^T = A^T\mathrm{Diag}(x)A\in\S_+^n$. Consequently, the OEDP problem with the information function $\phi$ can be formulated as the following mixed-integer optimization problem:
\begin{equation}
    \label{eq:oedp}\tag{OD}
    \min_{x\in\mathbb{R}^m}\; f(x):=-\log\,\phi(X(x))\quad \mathrm{s.t.}\quad \sum_{i = 1}^mx_i = N,\; x\in \Z_+^m.
\end{equation}
In the literature, the information function $\phi$ is frequently chosen as the matrix mean function of order $p\in\R$ \cite{kiefer1974general}:
\[
    \phi_p(X):= \begin{cases}
            \lambda_{\rm max}(X) & \textrm{if $p = \infty$} \\
            \left(\frac{1}{n}\mathrm{Tr}(X^p)\right)^{1/p} & \textrm{if $p\neq 0,\pm\infty$} \\
            \det(X)^{1/n} & \textrm{if $p = 0$} \\
            \lambda_{\rm min}(X) & \textrm{if $p = -\infty$} \\
            0 & \textrm{if $p \in  [-\infty, 0]$ and $X$ is singular}
        \end{cases},\quad \forall X\in \S^n_+,
\]
where $\lambda_{\rm min}(X)$ and $\lambda_{\rm max}(X)$ denote the minimal and maximal eigenvalues of $X$, respectively, $\mathrm{Tr}(\cdot)$ is the trace operator, and $\det(\cdot)$ denotes the determinant function. Particularly, the cases for $p = 0$ and $p=-1$ refer to D-optimal experimental design and A-optimal experimental design, respectively, which are two most important OEDP models in the literature \cite{pukelsheim2006optimal} and hence our representative examples in this paper. Note that for both criteria, the associated objective functions $f$ belong to the class of  self-concordant function \cite{sun2019generalized}, which is an important class of convex functions. In light of this, we assume for the rest of this paper that the objective function $f$ is  self-concordant.

\subsection{Related work}

As a mixed-integer optimization problem, \eqref{eq:oedp} is typically NP-hard (see, e.g., \cite{welch1982branch, nikolov2022proportional}), making the computation of its exact optimal solution extremely challenging, even when the objective function is convex. 

\textbf{Continuous relaxations.} To get rid of the integer constraints, one natural idea is to consider the continuous relaxation of \eqref{eq:oedp}, i.e., replace the constraint $x\in\Z_+^m$ with the nonnegative constraint $x\in\R_+^m$. Consequently, when $f$ is convex, one can apply well-developed and efficient convex optimization algorithms for solving the resulting relaxation problem. Particularly, convex optimization algorithms for A-optimal and D-optimal experimental designs include the multiplicative algorithm \cite{silvey1978algorithm, torsney1983moment, harman2009approximate}, the vertex direction method \cite{fedorov2013theory}, the vertex exchange method \cite{bohning1986vertex}, the cocktail algorithm \cite{yu2011d}, the interior point method \cite{lu2010interior}, the Frank-Wolfe method and its variants \cite{atwood1973sequences,lacoste2015global,ahipacsaouglu2015first}, projected and/or proximal Newton type methods \cite{tran2022new, liu2022newton, liang2024vertex}, and algorithms for solving conic programming formulations and/or relaxations \cite{vandenberghe1998determinant, sagnol2011computing, filova2012computing, toh2012implementation, papp2012optimal, ye2017computing, duarte2018adaptive}, to mention just a few. 

\textbf{Exact optimal design.} Though typically easier to solve, simply replacing the integer constraint with the nonnegative constraint can lead to approximate solutions that are different from the exact ones. In view of this, computing the exact solution of the mixed-integer optimization problem \eqref{eq:oedp} has long been an active research theme. The classical methods for solving the challenging optimization problem are mainly based on carefully designed heuristics approaches such as exchange methods \cite{atkinson2007optimum, mitchell2000algorithm}, rounding methods \cite{pukelsheim1992efficient, imhof2001efficiencies, pukelsheim2006optimal}, simulated annealing algorithms \cite{haines1987application, angelis2001optimal}, and genetic algorithms \cite{heredia2003genetic, limmun2013using}, without convergence guarantee. To compute provably exact optimal designs, the branch-and-bound (BnB) method represents one of the most popular approaches in the literature \cite{welch1982branch, ahipacsaouglu2021branch}. Recently, combining the BnB method with Frank-Wolfe methods \cite{hendrych2022convex, hendrych2023solving} has shown promising practical performance when compared with classical BnB approaches, given the fact that the underlying linear oracles can be evaluated efficiently by existing mixed-integer linear programming solvers such as \texttt{SCIP} \cite{bestuzheva2021scip}. Last but not least, reformulating the mixed-integer problem as a mixed-integer conic programming problem \cite{sagnol2015computing, kronqvist2019review, coey2020outer, duarte2023exact, harman2025mixed} provides alternating approaches for which efficient algorithms for solving conic programming can be fully explored.

\subsection{Contributions}
In this paper, we shall continue the research theme on BnB method for computing the exact experimental design. To this end, we adopt the BnB framework proposed in \cite{ahipacsaouglu2021branch} whose \texttt{Julia} implementation was recently provided by \cite{hendrych2023solving} based on the \texttt{Bonono.jl} package \footnote{Available at: \url{https://github.com/Wikunia/Bonobo.jl}.}. From the computational results provided in \cite{hendrych2023solving}, the BnB method of  \cite{ahipacsaouglu2021branch} demonstrates relative poor performance. The reasons are tow-fold: (1) The huge number of nodes to explore. (2) The excessive computational time for evaluating each node. Note that the first issue may be addressed by designing more effective and comprehensive branching, searching and pruning strategies. However, the design of such strategies can be highly nontrivial and problem dependent. On the other hand, the second issue can be effectively addressed if one can significantly reduce the time needed for evaluating nodes, all sharing a similar structure. In view of this, we focus on improving the efficiency of the BnB method via proposing a novel framework for accelerating the evaluation time for each node. Particularly, our approaches and contributions can be summarized as follows.
\begin{itemize}
    \item Instead of evaluating the nodes in the BnB method via the vertex exchange method of \cite{ahipacsaouglu2021branch}, we propose to apply the projected Newton method of \cite{liu2022newton} for solving the continuous convex constrained optimization with a generalized self-concordant objective on each node. The essential idea of the projected Newton method is to solve a sequence of convex quadratic programming (QP) approximations iteratively. Thanks to the elegant convergence properties developed in \cite{liu2022newton}, the number of QP approximations to be solved on each node is often small. This leads to significant reduction of the computational time, since the main computational bottleneck for evaluating a node in the BnB method falls into the computations or the updates of the objective function and its gradient. However, the vertex exchange method described in \cite{ahipacsaouglu2021branch} often requires significantly more iterations or potentially fails to converge altogether, since the global convergence of the method in \cite{ahipacsaouglu2021branch} remains unclear. Encouraging, a recent work \cite{liang2024vertex} has established the global convergence guarantees and demonstrated promising practical efficiency for the vertex exchange method when applied to the QP approximation as just mentioned. Consequently, we obtain a novel and efficiency framework with convergence guarantees for evaluating the nodes by combining the vertex exchange method with the projected Newton method.
    \item We develop \texttt{PNOD.jl}, an open-source \texttt{Julia} package, that can be applied to compute the exact D-optimal and A-optimal experimental designs. Our numerical studies validate the excellent efficiency of the proposed framework by showing that much more nodes can be explored in the BnB method within a given time limit, increasing the chance of obtaining the exact global optimum. We also highlight that though we use D-optimal and A-optimal designs as our illustrative examples, our package can be easily extended to other criteria as long as the objective function $f$ in \eqref{eq:oedp} is generalized self-concordant.
\end{itemize}

\subsection{Layout}
The rest of this paper is organized as follows. In Section \ref{sec:bnb}, we overview the generic BnB method that is applicable to general constrained optimization problems. In the same section, we also demonstrate how one can apply the same idea for computing the exact optimal experimental design. We then address the continuous relaxations in the BnB search tree in Section \ref{sec:pn}. In particular, we describe and compare two approaches, namely the vertex exchange method that is applied directly and the projected Newton method combining with the vertex exchange method. We conduct a set of extensive experiments in Section \ref{sec:exp} to validate the efficiency of the proposed framework via comparing with state-of-the-art methods. Finally, we conclude our paper in Section \ref{sec:conclusions}.

\section{The branch and bound method}\label{sec:bnb}
The branch-and-bound (BnB) method, first proposed by \cite{land2010automatic}, is recognized as one of the most fundamental and widely-used methodologies for computing exact global solutions to NP-hard discrete optimization problems; see \cite{morrison2016branch} for a comprehensive survey. To explain the main idea of the BnB method, let us consider the following general constrained optimization problem:
\begin{equation}
    \label{eq:gopt}
    \min_{x\in \cX}\; f(x),
\end{equation}
where $f:\cX\to\R$ is the objective function (with slightly abuse of notation) and $\cX$ denotes the search space. We assume that the set $\cX$ and subsets of $\cX$ admit continuous relaxations, and a feasible solution $\hat{x}\in \cX$ is given and saved globally. In the literature, the point $\hat{x}$ is often called the incumbent solution.

The high level idea of the generic BnB method is to implicitly enumerates all possible solutions of the interested problem using a tree structure, each node of which stores a subproblem. The BnB method typically requires the following three main strategies:
\begin{itemize}
    \item \textbf{Searching strategy:} the strategy that chooses the order to explore the nodes in the tree. 
    \item \textbf{Branching strategy:} the strategy guides how the search space is partitioned to generate new subproblem at a given node in the tree.
    \item \textbf{Pruning strategy:} the strategy that prevents the exploration of descendants of a given node that are guaranteed to be suboptimal.
\end{itemize}
At each iteration of the BnB method, the searching strategy selects a node in the search tree which contains a subproblem of the form $\min\{f(x)\;:\; x\in \cS\}$, where $\cS\subseteq \cX$. If the solution at this node, denoted as $x'\in \cS$, is feasible to \eqref{eq:gopt} and admits a better objective value than the globally saved incumbent objective (i.e., $x'\in \cX$ and $f(x') < f(\hat x)$), then the globally saved incumbent solution is replaced with $\hat x \leftarrow x'$. Otherwise, the algorithm chooses one of the following two actions.
\begin{itemize}
    \item The branching strategy partitions the current search space for the subproblem into a set of several new subproblems with search spaces $\cS_1,\dots, \cS_r$. Each newly generated subproblem will be stored at a new node that will be inserted into the search tree as a descendant of the current node. 
    \item If the pruning strategy suggests that no solution in $\cS$ has a better objective value than the global incumbent objective $f(\hat x)$, then the current node is pruned (or fathomed) and no descendant of the node will be generated (i.e., the branching strategy is not executed at this node).
\end{itemize}
Eventually, when all the nodes in the search tree are explored, the BnB method returns the globally saved incumbent solution $\hat x$. Clearly, $\hat x$ is the global optimal solution for \eqref{eq:gopt}. Based on the above description, the template of the generic BnB method is presented in Algorithm \ref{alg:bnb}. 

\begin{algorithm}[htb!]
\caption{The generic BnB method.}\label{alg:bnb}
\begin{algorithmic}[1]
\State Generate and save an incumbent solution $\hat{x}$.
\State Set $\cT = \{(\cX, x(\cX))\}$ where $x(\cX)$ denote the optimal solution of the continuous relaxation problem.
\While{$\cT \neq \emptyset$}
\State Select a subproblem $(\cS, x(\cS))$ from $\cT$ based on the search strategy.
\If{$x(\cS)\in \cX$ and $f(x(\cS)) < f(\hat{x})$}
    \State Set $\hat {x} = x(\cS)$.
\EndIf
\If{$\cS$ can not be pruned by the pruning strategy}
    \State Partition $\cS$ into $\cS_1,\dots, \cS_r$ based on the branching strategy.
    \State For each $i$, solve the continuous relaxation problem associated with $\cS_i$ to get $x(\cS_i)$.
    \State Insert $(\cS_1, x(\cS_1)),\dots, (\cS_r, x(\cS_r))$ into $\cT$ as new subproblems.
\EndIf
\State Remove $(\cS, x(\cS))$ from $\cT$.
\EndWhile
\State \textbf{Output:} the incumbent solution $\hat{x}$.
\end{algorithmic}
\end{algorithm}

Next, we present the BnB method proposed in previous works \cite{ ahipacsaouglu2021branch}, which is a direct application of the generic BnB method just introduced, for computing the exact optimal design of \eqref{eq:oedp}. Clearly, in our case $f(x):= -\log\phi(X(x))$ and $\cX:=\{x\in\R^m\;:\; \sum_{i = 1}^mx_i = N,\; x\in \Z_+^m\}$. 

Obtaining a feasible incumbent solution $\hat{x}$ turns out to be relatively easy by applying a simple rounding heuristic, as suggested in \cite{ahipacsaouglu2021branch, hendrych2023solving}. First, we start with the zero vector $\hat{x}$ and choose an upper bound vector $\hat{u}\in \Z_{++}^m$ such that $\sum_{i = 1}^m\hat{u}_i \geq N$. Then, find an index set $J$ that corresponds to a set of $n$ linear independent rows of $A\in\R^{m\times n}$ with full column rank and set $\hat{x}_J = \hat{u}_J$. If the resulting vector $\hat{x}$ is not feasible, then we add certain experiments to the design point corresponding to zero experiments, if the total number of experiments in $\hat{x}$ is less than $N$. Otherwise, take away one experiment from the design point $\hat{x}$ with most number of experiments. Repeat the process until a feasible incumbent solution $\hat{x}$ is found. The detailed procedure of the rounding heuristic is described in Algorithm \ref{alg:round}.

\begin{algorithm}[htb!]
\caption{The rounding heuristic.}\label{alg:round}
\begin{algorithmic}[1]
\State \textbf{Input:} matrix $A\in\mathbb{R}^{m\times n}$ with $n < m$ and full column rank, an upper bound vector $\hat {u}\in \Z^m_{++}$ such that $\sum_{i=1}^m\hat{u}_i \geq N$.
\State Set $\hat{x} = 0$ as the zero vector in $\R^m$.
\State Find a set of $n$ linearly independent rows of $A$ and record their indices as $J$.
\State Set $\hat{x}_J = \hat{u}_J$.
\If{$\sum_{i = 1}^m \hat{x}_i > N$}
    \While{$\sum_{i = 1}^m \hat{x}_i > N$}
        \State Find $j_{\rm max}\in \mathrm{argmax}\{\hat{x}_j\;:\; j\in J\}$.
        \State Set $\hat{x}_{j_{\rm max}} = \hat{x}_{j_{\rm max}}-1$.
    \EndWhile
\ElsIf{$\sum_{i = 1}^m \hat{x}_i < N$}
    \While{$\sum_{i = 1}^m \hat{x}_i < N$}
        \State Find an integer $j$  not in $J$.
        \State Set $\hat{x}_{j} = \min\{N-\sum_{i=1}^m \hat{x}_i, \hat{u}_{j}\}$. 
        \State Set $J = J\cup \{j\}$.
    \EndWhile
\EndIf 
\State \textbf{Output:} $\hat{x}$.
\end{algorithmic}
\end{algorithm}

\begin{remark}
    \label{remark:initial}
    Note that the generated point $\hat{x}$ after executing lines 2 -- 4 in Algorithm \ref{alg:round} can be replaced by other candidates. For example, one may solve the following continuous convex optimization problem:
    \[
        \min_{x}\; f(x)\quad \mathrm{s.t.}\quad \sum_{i=1}^mx_i = N,\; 0\leq x_i \leq \hat{u}_i,\; \forall i = 1,\dots, n,
    \]
    to get the optimal solution $\tilde{x}$. Then, $\hat{x}$ is set to be the vector obtaining from rounding the vector $\tilde{x}$. However, based on our numerical experience, there is no significant difference in practical performance.
\end{remark}

Since we use the BnB framework implemented by the \texttt{Julia} package \texttt{Bonobo.jl}, the search strategy, branching strategy and pruning strategy are restricted to those being implemented in the package. Specifically, we use the best first search (BFS) strategy, which traverses $\cT$ via picking the node/subproblem with the lowest bound first. Here, for any subproblem $(\cS, x(\cS))\in \cT$, the bound refers to $f(x(\cS))$. For the branching strategy, we adopt the binary branching strategy, which selects the branching variable that is furthest away from being an integer. More precisely, suppose that $\cS:=\{x\in\R^m\,:\, \sum_{i=1}^mx_i = N,\; \ell \leq x\leq u,\; x\in\Z_+^m\}$ denote the search space for the selected subproblem and $x(\cS)$ denotes the optimal solution for the continuous relaxation problem $\min\{f(x)\,:\, \sum_{i=1}^mx_i = N,\; \ell_i \leq x\leq u_i\}$, where $\ell, u\in \R^m$ satisfy $0\leq \ell\leq u$. Then, the binary branching strategy selects an index $j\in \mathrm{argmax}\{|x(\cS)_i-[x(\cS)_i]|\,:\, i = 1,\dots, m\}$ and partitions the search space $\cS$ into the following two subsets of $\cS$:
\begin{equation}
\label{eq:branchingS}
    \begin{aligned}
    \cS_1 := &\; \left\{x\in\R^m\,:\, \sum_{i=1}^mx_i = N,\; \ell_i \leq x_i \leq u_i,\;\forall i\neq j, \, \ell_j \leq x_j \leq \lfloor x(S)_j \rfloor,\, x\in\Z_+^m\right\}, \\
    \cS_2 := &\; \left\{x\in\R^m\,:\, \sum_{i=1}^mx_i = N,\; \ell_i \leq x_i \leq u_i,\;\forall i\neq j, \, \lceil x(S)_j \rceil \leq x_j \leq u_j,\, x\in\Z_+^m\right\}.
\end{aligned}
\end{equation}
Upon obtaining $\cS_1$ and $\cS_2$, we shall next solve two continuous relaxation problems to get $x(\cS_1)$ and $x(\cS_2)$, i.e., 
\begin{equation}
    \label{eq:evaluatenode}
    \begin{aligned}
        x(\cS_1):= &\; \mathrm{argmin}\left\{f(x)\,:\, \sum_{i=1}^mx_i = N,\; \ell_i \leq x_i \leq u_i,\;\forall i\neq j, \, \ell_j \leq x_j \leq \lfloor x(S)_j \rfloor\right\}, \\
         x(\cS_2):= &\; \mathrm{argmin}\left\{f(x)\,:\, \sum_{i=1}^mx_i = N,\; \ell_i \leq x_i \leq u_i,\;\forall i\neq j, \, \lceil x(S)_j \rceil \leq x_j \leq u_j\right\}.
    \end{aligned}
\end{equation}
Finally, we choose the most commonly use pruning strategy based on lower bounds associated with subproblems in $\cT$. It is clear that for a chosen subproblem $(\cS, x(\cS))$, if $f(x(\cS)) > f(\hat{x})$, then for any feasible point $x\in \cS$, $f(x) \geq f(x(\cS)) > f(\hat{x})$ and hence it is safe to prune this subproblem. Moreover, in real implementation, it is a common practice to set up certain termination conditions so that the user can gain more flexibilities when running the algorithm. In our implementation, we terminate the algorithm if one of the following termination conditions is met:
\begin{itemize}
    \item $\cT = \emptyset$;
    \item A time limit is reached;
    \item The absolute gap $|f(\hat{x}) - f(x(\cS))|$ is less than a given tolerance \texttt{abstol};
    \item The relative gap $|f(\hat{x}) - f(x(\cS))|/\min\{|f(\hat{x})|, |f(x(\cS))|\}$ is less than a given tolerance \texttt{reltol}.
\end{itemize}

With all the previous detailed discussions, we are now able to present the pseudocode for the BnB method for solving the problem \eqref{eq:oedp} in Algorithm \ref{alg:bnbod}.

\begin{algorithm}[htb!]
\caption{The BnB method for exact optimal experimental design.}\label{alg:bnbod}
\begin{algorithmic}[1]
\State Apply Algorithm \ref{alg:round} for generating an incumbent solution $\hat{x}$.
\State Set $\cX:=\{x\in\R^m\;:\; \sum_{i = 1}^mx_i = N,\; x\in \Z_+^m\}$.
\State Compute $x(\cX) = \min\{f(x)\,:\, \sum_{i = 1}^mx_i = N, x\in \R_+^m\}$.
\State Set $\cT = \{(\cX, x(\cX)\}$.
\While{a termination condition is not met}
\State Select a subproblem $(\cS, x(\cS))$ from $\cT$ with the lowest bound $f(x(S))$.
\If{$x(\cS)\in \cX$ and $f(x(\cS)) < f(\hat{x})$}
    \State Set $\hat {x} = x(\cS)$.
\EndIf
\If{$f(x(\cS)) \leq f(\hat{x})$}
    \State Partition $\cS$ into $\cS_1$ and $\cS_2$ as in \eqref{eq:branchingS}.
    \State Compute $x(\cS_1)$ and $x(\cS_2)$ as in \eqref{eq:evaluatenode}.
    \State Insert $(\cS_1, x(\cS_1)), (\cS_2, x(\cS_2))$ into $\cT$ as new subproblems.
\EndIf
\State Remove $(\cS, x(\cS))$ from $\cT$.
\EndWhile
\State \textbf{Output:} the incumbent solution $\hat{x}$.
\end{algorithmic}
\end{algorithm}

\section{Fast computation of continuous relaxations}\label{sec:pn}

We see that the practical efficiency of Algorithm \ref{alg:bnbod} depends on the choices of search strategy, branching strategy, termination conditions, and the efficiency in solving a continuous optimization problem of the form:
\begin{equation}
    \label{eq:cvx}
    \min_{x\in \R^m} \; f(x)\quad \mathrm{s.t.}\quad e^Tx = N, \; \ell\leq x\leq u,
\end{equation}
where $e\in\R^m$ denotes the vector of all ones, $0\leq \ell \leq u \leq Ne$ are two given vectors. In this section, we focus on solving the above problem efficiently. Without loss of generality, we assume for the rest of this section that the feasible set of \eqref{eq:cvx}, denoted as $\cF:=\{x\in\R^m\;:\; e^Tx = N, \ell\leq x\leq u\}$, is nonempty, i.e., $e^T\ell\leq N \leq e^TN$. Clearly, $\cF$ is compact and the problem \eqref{eq:cvx} admits an optimal solution $x^*$.

Before delving into the detailed algorithmic descriptions, we shall provide the expressions of the gradient and Hessian of the objective functions associated with A-optimal and D-optimal experimental designs, respectively, as they are needed in the practical implementation. 

\begin{example}[A-optimal experimental design]
    \label{eg-A-design}
    The objective function associated with A-criterion (ignoring the constant scaling factor  for simplicity) is defined as 
    \[
        f(x) = \log\mathrm{tr} \;\left(\sum_{i = 1}^mx_ia_ia_i^T\right)^{-1},\quad x\in\mathrm{dom}(f).
    \]
    By the definition of the first-order partial derivatives, it is straightforward to verify that 
    \begin{align*}
    &\; \frac{\partial f(x)}{\partial x_j} \\
    = &\; \lim_{t\to 0}\frac{f(x+te_j) - f(x)}{t} \\
    = &\; \lim_{t\to 0}\frac{1}{t}\log\left(\frac{\mathrm{tr} \;\left(\left(\sum_{i = 1}^mx_ia_ia_i^T\right)^{-1} - \left(\sum_{i = 1}^mx_ia_ia_i^T\right)^{-1}\left(ta_ja_j^T\right)\left(\sum_{i = 1}^mx_ia_ia_i^T\right)^{-1} + O(t^2)I\right)}{\mathrm{tr} \;\left(\sum_{i = 1}^mx_ia_ia_i^T \right)^{-1}}\right) \\
    = &\; -\frac{\mathrm{tr} \;\left( \left(\sum_{i = 1}^mx_ia_ia_i^T\right)^{-1}\left(a_ja_j^T\right)\left(\sum_{i = 1}^mx_ia_ia_i^T\right)^{-1} \right)}{\mathrm{tr} \;\left(\sum_{i = 1}^mx_ia_ia_i^T \right)^{-1}} \\
    = &\; -\frac{ a_j^T\left(\sum_{i = 1}^mx_ia_ia_i^T\right)^{-2} a_j}{\mathrm{tr} \;\left(\sum_{i = 1}^mx_ia_ia_i^T \right)^{-1}},\quad \forall j = 1,\dots, m,
\end{align*}
where the second equality is derived from \cite[Ch. 9, Sec. 14]{magnus2019matrix}. Consequently, it holds that 
\[
    \nabla f(x) = -\frac{1}{\mathrm{tr}(A^T\mathrm{Diag}(x)A)^{-1}}\mathrm{diag}\left(A(A^T\mathrm{Diag}(x)A)^{-2}A^T\right),\quad x\in\mathrm{dom}(f).
\]
Similar reason shows that 
\begin{align*}
    \frac{\partial^2 f(x)}{\partial x_j\partial x_k} 
    =  &\; -\frac{\frac{\partial\left(a_j^T\left(\sum_{i = 1}^mx_ia_ia_i^T\right)^{-2} a_j\right)}{\partial x_k}}{\mathrm{tr}(A^T\mathrm{Diag}(x)A)^{-1}} + \frac{\frac{\partial\left(\mathrm{tr} \;\left(\sum_{i = 1}^mx_ia_ia_i^T \right)^{-1}\right)}{\partial x_k}a_j^T\left(\sum_{i = 1}^mx_ia_ia_i^T\right)^{-2} a_j}{\left(\mathrm{tr}(A^T\mathrm{Diag}(x)A)^{-1}\right)^2} \\
    = &\; \frac{2a_j^T\left(\sum_{i = 1}^mx_ia_ia_i^T\right)^{-2} a_ka_k^T\left(\sum_{i = 1}^mx_ia_ia_i^T\right)^{-1}a_j}{\mathrm{tr} \;\left(\sum_{i = 1}^mx_ia_ia_i^T \right)^{-1}} \\ 
    &\; - \frac{a_k^T\left(\sum_{i = 1}^mx_ia_ia_i^T \right)^{-2}a_ka_j^T\left(\sum_{i = 1}^mx_ia_ia_i^T\right)^{-2} a_j}{\left(\mathrm{tr} \;\left(\sum_{i = 1}^mx_ia_ia_i^T \right)^{-1}\right)^2},\quad \forall \; 1\leq j, k \leq m.
\end{align*}
Therefore, we see that 
\begin{align*}
    \nabla^2 f(x) = &\; \frac{2}{\mathrm{tr}(A^T\mathrm{Diag}(x)A)^{-1}}\left(A(A^T\mathrm{Diag}(x)A)^{-2}A^T\right) \circ \left(A(A^T\mathrm{Diag}(x)A)^{-1}A^T\right) \\
    &\; -\frac{1}{\left(\mathrm{tr}(A^T\mathrm{Diag}(x)A)^{-1}\right)^2}\mathrm{diag}\left(A(A^T\mathrm{Diag}(x)A)^{-2}A^T\right) \mathrm{diag}\left(A(A^T\mathrm{Diag}(x)A)^{-2}A^T\right)^T.
\end{align*}

\end{example}

\begin{example}[D-optimal experimental design]
    \label{eg-D-design}
    The objective function associated with D-criterion (ignoring the constant scaling factor for simplicity) is defined as 
\[
    f(x) = -\log\det\left(\sum_{i = 1}^mx_ia_ia_i^T\right),\quad x\in\mathrm{dom}(f).
\]
Then we see that 
\begin{align*}
    \frac{\partial f(x)}{\partial x_j} = &\; \lim_{t\to0}\frac{1}{t}(f(x+te_j) - f(x)) \\
    = &\; \lim_{t\to0}\frac{1}{t}\left(-\log\det\left(\sum_{i = 1}^mx_ia_ia_i^T\right)\left(I + \left(\sum_{i = 1}^mx_ia_ia_i^T\right)^{-1/2}(ta_ja_j^T)\left(\sum_{i = 1}^mx_ia_ia_i^T\right)^{-1/2}\right) \right.\\
    &\; + \left. \log\det\left(\sum_{i = 1}^mx_ia_ia_i^T\right)\right) \quad \textrm{(By \cite[Ch. 9, Sec. 14]{magnus2019matrix})} \\
    = &\; \lim_{t\to0}-\frac{1}{t}\log\det \left(I + \left(\sum_{i = 1}^mx_ia_ia_i^T\right)^{-1/2}(ta_ja_j^T)\left(\sum_{i = 1}^mx_ia_ia_i^T\right)^{-1/2}\right).
\end{align*}
Consider the following spectral decomposition 
\[
    \left(\sum_{i = 1}^mx_ia_ia_i^T\right)^{-1/2}(a_ja_j^T)\left(\sum_{i = 1}^mx_ia_ia_i^T\right)^{-1/2} = Q\Lambda Q^T,\quad \Lambda = \mathrm{Diag}(\lambda_1,\dots,\lambda_m).
\]
Then, we see that 
\begin{align*}
    \frac{\partial f(x)}{\partial x_j} = \lim_{t\to 0} -\frac{1}{t}\sum_{i = 1}^m\log (1+t\lambda_i) 
    =  -\sum_{i = 1}^m\lambda_i 
    =   -a_j^T\left(\sum_{i = 1}^mx_ia_ia_i^T\right)^{-1}a_j.
\end{align*}
Clearly, we have 
\[
\nabla f(x) = -\mathrm{diag}(A(A^T\mathrm{Diag}(x)A)^{-1}A^T),\quad x\in\mathrm{dom}(f).
\]
Similarly, we see that 
\begin{align*}
    \frac{\partial^2 f(x)}{\partial x_j\partial x_k} = &\; -\frac{\partial \left(a_j^T\left(\sum_{i = 1}^mx_ia_ia_i^T\right)^{-1}a_j\right)}{\partial x_k} \\
    = &\; -\lim_{t\to0}\frac{1}{t}\left(a_j^T\left(\left(\sum_{i = 1}^mx_ia_ia_i^T\right)^{-1} - t\left(\sum_{i = 1}^mx_ia_ia_i^T\right)^{-1}(a_ka_k^T) \left(\sum_{i = 1}^mx_ia_ia_i^T\right)^{-1}\right)a_j \right. \\
    &\; - \left. a_j^T\left(\sum_{i = 1}^mx_ia_ia_i^T\right)^{-1}a_j\right) \\
    = &\; \left( a_j^T\left(\sum_{i = 1}^mx_ia_ia_i^T\right)^{-1}a_k\right)^2.
\end{align*}
Hence, it holds that 
\[
    \nabla^2f(x) = (A(A^T\mathrm{Diag}(x)A)^{-1}A^T)\circ (A(A^T\mathrm{Diag}(x)A)^{-1}A^T),\quad x\in\mathrm{dom}(f).
\]
\end{example}

\subsection{The vertex exchange method for \eqref{eq:cvx}}

For the A-optimal and D-optimal design problem, the associated continuous relaxation is solved via a vertex exchange method in \cite{ahipacsaouglu2021branch} which is a direct extension of the method proposed in \cite{bohning1986vertex} for D-optimal design. The key idea of the vertex exchange method is to quickly generate a descent direction of the objective based on the entries of its gradient with respect to inactive bound constraints. The idea is summarized in the following lemma.
\begin{lemma}[{\cite[Lemma 4.1]{ahipacsaouglu2021branch}}]
\label{lemma:vem}
    Let $x\in \cF$ be a feasible solution and also 
    \begin{equation}
        \label{eq:jk}
        j \in \mathrm{argmin}\{\nabla f(x)_i\;:\; x_i < u_i\},\quad k \in \mathrm{argmax}\{\nabla f(x)_i\;:\; x_i > \ell_i\}.
    \end{equation}
    If $\nabla f(x)_j \geq \nabla f(x)_k$, then $x$ is an optimal solution for \eqref{eq:cvx}. Otherwise, $d:= e_j - e_k$ is a feasible descent direction of $f$, i.e., $\nabla f(x) ^Td < 0$,
    where $e_i\in \R^m$ denotes the $i$-th column of the identity matrix of size $m\times m$ for $1\leq i\leq m$. 
\end{lemma}
Lemma \ref{lemma:vem} also suggests an optimality condition for the problem \eqref{eq:cvx}, i.e., $\nabla f(x)_j \geq \nabla f(x)_k$. This condition can be used to define a termination criterion in practical implementation, since one can only afford to solve the problem approximately. From the lemma, we also see that for the generated descent direction, namely $d$, there exists a scalar $\eta > 0$ such that the new point defined as 
\begin{equation}
    \label{eq:vem-eta}
    x^+ = x + \eta e_j - \eta e_k
\end{equation}
provides a smaller objective value, i.e., $f(x^+) < f(x)$. To make $f(x^+)$ as small as possible while keeping $x^*$ being feasible, the optimal $\eta$ should be computed as follows:
\begin{equation}
    \label{eq:vem-opt-eta}
    \min_{\eta\in\R}\; f(x+\eta e_j - \eta e_k)\quad \mathrm{s.t.}\quad 0\leq \eta \leq u_j - x_j,\; 0\leq \eta \leq x_k - \ell_k.
\end{equation}
It is shown in \cite{ahipacsaouglu2021branch} that the optimal step size for the A-criterion or the D-criterion admits an analytical expression. To summarize, the direct application of the vertex exchange method for solving problem \eqref{eq:cvx} is presented in Algorithm \ref{alg:vem-direct}.

\begin{algorithm}[htb!]
\caption{The vertex exchange method for \eqref{eq:cvx} \cite{ahipacsaouglu2021branch}.}\label{alg:vem-direct}
\begin{algorithmic}[1]
\State \textbf{Input:} A feasible point $x\in\cF$ and a termination tolerance \texttt{tol}.
\State Calculate $\nabla f(x)$.
\State Find the indices $j$ and $k$ by \eqref{eq:jk}.
\While{$\nabla f(x)_k / \nabla f(x)_j -1 > \texttt{tol}$}
    \State Compute the step size $\eta$ by \eqref{eq:vem-opt-eta}.
    \State Update $x$ by \eqref{eq:vem-eta}.
    \State Update the gradient $\nabla f(x)$.
    \State Find the indices $j$ and $k$ by \eqref{eq:jk}.
\EndWhile
\State \textbf{Output:} $x$.
\end{algorithmic}
\end{algorithm}

Theoretically and numerically, the computational bottleneck of the Algorithm \ref{alg:vem-direct} lies in updating the gradient, which requires recomputing the inverse of the information matrix $X(x)$ directly or implicitly through advanced numerical techniques such as low-rank Cholesky updates and downdates \cite{seeger2004low, bartels1989cholesky}. Moreover, the global convergence of Algorithm \ref{alg:vem-direct} remains unclear. Though our numerical experience suggests that the algorithm is empirically convergence, the convergence rate of Algorithm \ref{alg:vem-direct} is typically slow. Consequently, Algorithm \ref{alg:vem-direct} shows poor efficiency in practice, as also indicated by \cite{hendrych2023solving}. 
In view of this, the rest of this section is dedicated to the design of a convergent and efficient algorithmic framework for solving \eqref{eq:cvx} that only requires a few computations of the gradient $\nabla f(\cdot)$. 

\subsection{The projected Newton method for \eqref{eq:cvx}}
As mentioned in the introduction, we assume that $f:\R^m\to \R$ is  self-concordant. The definition of the self-concordance is given explicitly in the following definition to make the paper self-contained. Readers are referred to \cite{sun2019generalized} for a comprehensive treatment of the concept together with fruitful examples arising from a wide range of practical applications.

\begin{definition}
    \label{def:selfconcordant}
    A three-times differentiable, convex function $f:\R^m\to \R\cup\{\pm\infty\}$ is said to be self-concordant if there exists a constant $M_f\geq 0$ such that the inequality 
    \[
        \left\lvert \nabla^3f(x)[u, u, u] \right\rvert \leq M_f \norm{u}_x^{3/2},\quad \forall\; u\in\R^m,
    \]
    holds for any $x\in \mathrm{dom}(f):= \{x\in\R^m\;:\; f(x)<\infty\}$, where $\norm{u}_x:= \sqrt{\inprod{u}{\nabla^2 f(x) u}}$ denotes the local norm of $u$ associated with $f$ at $x$, and 
    \[
        \nabla^3f(x)[u, u, u]:= \lim_{t\to 0}\frac{1}{t}(\nabla^2 f(x+tu) - \nabla^2 f(x)).
    \]
    If $M_f=2$, then we say that $f$ is standard self-concordant.
\end{definition}
\cite[Corollary 4.1.2]{nesterov2013introductory} suggests that any self-concordant function $f$ can be rescaled to the standard form by considering $\hat{f}:= \frac{M_f}{4}f$. Moreover, the function $f$ associated with A-criterion or D-criterion is standard self-concordant on the feasible set $\cF$ with a proper rescaling \cite{nesterov1994interior, hendrych2023solving}. The following lemma gives a sufficient condition for the nonsingularity of the Hessian $\nabla^2 f(\cdot)$, which apparently applies to $f$ under the A-criterion and D-criterion.
\begin{lemma}[{\cite[Theorem 4.1.3]{nesterov2013introductory}}]
\label{lemma-noline}
    Let $f$ be self-concordant. If $\mathrm{dom}(f)$ contains no straight line, then the Hessian $\nabla^2 f(x)$ is positive definite at any $x$ from $\mathrm{dom}(f)$.
\end{lemma}

Since $f$ is self-concordant, it can be well-approximated by a quadratic surrogate function:
\[
    f(z)\approx q(z; x) := f(x) + \inprod{\nabla f(x)}{z - x} + \frac{1}{2}\inprod{z - x}{\nabla^2f(x)(z-x)},\quad \forall z\in \mathrm{dom}(f),
\]
where $x\in \mathrm{dom}(f)$ is given. Suppose that $\mathrm{dom}(f)$ contains no straight line, from Lemma \ref{lemma-noline} we see that $\nabla^2f(x)$ is positive definite. Consequently, the following strongly convex quadratic programming (QP) problem:
\begin{equation}
    \label{eq:qp}
    \min_{z\in\R^m}\; q(z;x)\quad \mathrm{s.t.}\quad z\in\cF,
\end{equation}
has a unique solution, denoted as $S(x)$. 

Using the above notation, the essential idea of the projected Newton method proposed in \cite{liu2022newton} for solving \eqref{eq:cvx} is to iteratively perform the following:
\begin{equation}
    \label{eq:pn}
    \hat{x} \approx S(x),\quad \hat{x}\in \cF,\quad x \leftarrow x + \eta (\hat{x} - x),
\end{equation}
where $\eta > 0$ is the step size. If the constraint $x \in \cF$ is omitted, then the scheme reduced to the classical Newton method, which is one of the most fundamental and efficient algorithms for solving convex optimization problems. Due to the presence of the constraint set $\cF$, the scheme is called the projected Newton (PN) method. Three key issues to be addressed in the (PN) method are given as follows:
\begin{itemize}
    \item How to solve the underlying QP problem \eqref{eq:qp} approximately and produce a point $\hat{x}\in\cF$?
    \item How to quantify and control the inexactness for computing the point $\hat{x}$?
    \item How to determine the step size $\eta$?
\end{itemize}
We assume the availability of a QP solver, with its design details addressed in the following subsection, and will concentrate on resolving the remaining two issues. 

To quantify the optimality of a candidate solution $x$, we rely on the following optimality condition for the QP problem \eqref{eq:qp}:
\begin{equation}
    \label{eq:opt-qp}
    \inprod{\nabla q(S(z);z)}{S(z) - x} \leq 0,\quad \forall\; x\in \cF.
\end{equation}
Then, we say that a point $S_\epsilon(z)\in\R^m$ is an $\epsilon$-solution of \eqref{eq:qp} if the following inequality holds:
\begin{equation}
    \label{eq:opt-eps}
    \max_{x\in\cF}\; \inprod{\nabla q(S_\epsilon(z);z)}{S_\epsilon(z) - x} \leq \epsilon^2.
\end{equation}
Clearly, the inexact tolerance $\epsilon$ should be non-increasing. For the calculation of the step size $\eta$, ideally we want $\eta = 1$ since it is desirable for Newton-type methods. However, a damped-step is needed at the early stage of the algorithm to guarantee the convergence. The full step size is only acceptable when an iterate kicks into the fast convergence region of the algorithm. For later usage, we also define the following function $h:\R_+ \to \R_+$ as 
\[
    h(\tau):= \frac{\tau(1 - 2\tau + 2\tau^2)}{(1-2\tau)(1-\tau)^2 - \tau^2},\quad \tau\in\R_+.
\]
With these preparations, we can present the PN method of \cite{liu2022newton} in Algorithm \ref{alg:pn}.

\begin{algorithm}[htb!]
\caption{The projected Newton method for \eqref{eq:cvx} \cite{liu2022newton}.}\label{alg:pn}
\begin{algorithmic}[1]
\State \textbf{Input:} A feasible initial point $x^0\in\cF\cap \mathrm{dom}(f)$, termination tolerance $\texttt{tol} > 0$, parameters $\sigma, \delta \in (0,1)$, $\beta\in (0, 0.5)$ and $C, C_1>1$.
\State Set $\lambda = \beta/\sigma$ and $\epsilon = \min\left\{\beta/C, C_1h^{-1}(\beta)\right\}$.
\State Set $x = x^0$.
\State Compute $\nabla f(x)$ and $\nabla^2 f(x)$.
\While{$\nabla f(x)_k / \nabla f(x)_j -1 > \texttt{tol}$}
    \State Find an approximate solution $S_\epsilon(x)\in \cF$ of the following QP:
    \[
        \min_{z\in\R^m}\; f(x) + \inprod{\nabla f(x)}{z-x} + \frac{1}{2}\inprod{z-x}{\nabla^2f(x)(z-x)}\quad \mathrm{s.t.}\quad z\in\cF
    \]
    \State Set $d = \hat{x} - x$ and $\gamma = \norm{d}_x$.
    \If{$\epsilon + \gamma \leq h^{-1}(\beta)$ or $\lambda\leq \beta$}
        \State Set $\lambda\leftarrow \sigma \lambda$ and $\epsilon \leftarrow \sigma\epsilon$.
        \State Set $\eta = 1$.
        \State Set $x\leftarrow x + \eta d$.
    \Else 
        \State Set $\eta = \delta (\gamma^2 - \epsilon^2) / (\gamma^3+\gamma^2 - \epsilon^2\gamma)$.
        \State Set $x\leftarrow x + \eta d$.
    \EndIf
    \State Compute $\nabla f(x)$ and $\nabla^2 f(x)$.
\EndWhile
\State \textbf{Output:} $x$.
\end{algorithmic}
\end{algorithm}

We see that the algorithm automatically adjust the inexactness tolerance and parameters used to determine the step size so that the following convergence properties can be established.

\begin{theorem}[{\cite[Theorem 1 \& 2]{liu2022newton}}]
    \label{thm-conv-pn}
    Suppose that $\cF$ is compact and $f:\R^m\to \R\cup\{\pm\infty\}$ is standard self-concordant whose effective domain $\mathrm{dom}(f)$ contains no straight line. Let $\omega(\tau):=\tau - \log(1+\tau)$. Then within at most $\left\lceil\frac{f(x^0) - f(x^*)}{\delta \omega\left(\frac{1-2C_1}{1-C_1}h^{-1}(\beta)\right)}\right\rceil$ iterations, we can guarantee that $\gamma + \epsilon\leq h^{-1}(\beta)$, which leads to full step size, i.e., $\eta = 1$ and implies that $\norm{x- x^*}_{x^*}\leq \beta$. 

    Suppose that the initial point satisfies $\norm{x^0 - x^*}_{x^*}\leq \beta$ and the triple $(\sigma, \beta, C)$ satisfies the following conditions: 
    \[
        \frac{1}{C(1-\beta)} + \frac{\beta}{(1-2\beta)(1-\beta)^2} \leq \sigma,\quad \frac{1}{C} + \frac{1}{1-2\beta} \leq 2,
    \]
    Then the full step size $\eta = 1$ is always acceptable and $\norm{x - x^*}_{x^*}$ converges to zero linearly. 
\end{theorem}
Theorem \ref{thm-conv-pn} essentially states that after a finite number of damped step (i.e., $\eta < 1$) PN iterations, the iterate $x$ will kick into the fast linear convergence region such that full step PN iterations are acceptable. Moreover, the linear convergence of Algorithm \ref{alg:pn} indicates that only a few number of iterations is needed, resulting in significant reduction in computational time for evaluating the gradient of $f$ when compared with Algorithm \ref{alg:vem-direct}. As a result, the PN method is more attractive than the vertex exchange method, since it admits appealing convergence properties. Careful readers will note that evaluating the Hessian $\nabla^2 f(x)$ is necessary at each iteration of the PN method. However, in practice, it is often possible to reuse the computations performed for evaluating the gradient $\nabla f(x)$, which can significantly reduce the computational cost of evaluating the Hessian.

\subsection{The vertex exchange method for \eqref{eq:qp}}

By applying the PN method as described in the previous subsection for the purpose of reducing the number of iterations, it is obvious that the practical efficiency of the PN method now depends on the efficiency in solving the QP problem of the form \eqref{eq:qp} at each iteration. Though the convergence of the vertex exchange method for solving \eqref{eq:cvx} remains unclear, a recent work \cite{liang2024vertex} has shown that applying the vertex exchange method for solving the QP problem \eqref{eq:qp} admits global convergence and excellent numerical performance when compared with state-of-the-art solvers. In view of this, the main purpose of this subsection is to describe the vertex exchange method for \eqref{eq:qp} and present its convergence properties. For the rest of this section, we assume without loss of generality that $e^T\ell < N < e^Tu$ and $\ell < u$ elementwise. Otherwise, if $e^T\ell = N$, then $x^* = \ell$, and if $e^Tu = N$, then $x^*=u$. Moreover, if there exists $i$ such that $\ell_i = u_i$, then $x^*_i = \ell_i = u_i$, then the QP problem can be reformulated as a smaller-scale problem that is of the same form of \eqref{eq:qp} and satisfies the above assumption. 

Similar to Lemma \ref{lemma:vem}, we can obtain a descent direction of $q(z;x)$ based on the information of its gradient with respect to inactive bound constraints. Indeed, we have the following lemma.

\begin{lemma}[{\cite[Lemma 3]{liang2024vertex}}]
    \label{lemma:vem-qp}
    Let $x,z\in \cF$ be two feasible points of \eqref{eq:qp} and select two indices 
    \begin{equation}
        \label{eq:vemqpjk}
        j\in \mathrm{argmin}\left\{\nabla q(z;x)_i\;:\; z_i < u_i\right\},\quad k\in \mathrm{argmax}\left\{\nabla q(z;x)_i\;:\; z_i > \ell_i\right\}.
    \end{equation}
    Then, $z$ is an optimal solution of \eqref{eq:qp} if and only $\nabla q(z;x)_k\leq \nabla q(z;x)_j$. Define a new point $z^+$ via:
    \begin{equation}
        \label{eq:qp-newz}
        z^+ := z + \eta (e_j - e_k),\quad \eta \in [0, \min\{u_j - z_j, z_k - \ell_k\}].
    \end{equation}
    Then, $z^+ \in \cF$, and if $\nabla q(z;x)_k > \nabla q(z;x)_j$, $d := e_j - e_k$ is a descent direction of $q(\cdot;x)$.
\end{lemma}

We see that, in order to make $q(z^+;x)$ as small as possible, the optimal step size $\eta$ can be computed by solving the following box-constraint optimization problem:
\[
    \min_{\eta\in\R}\; q(z+\eta(e_j-e_k);x)\quad \mathrm{s.t.}\quad \eta \in [0, \min\{u_j - z_j, z_k - \ell_k\}],
\]
which admits an analytical optimal solution (see, e.g., \cite[Lemma 4]{liang2024vertex}):
\begin{equation}
    \label{eq:etaqp}
    \eta = \min\left\{u_j - z_j, z_k - \ell_k, \frac{\nabla q(z;x)_k- \nabla q(z;x)_j}{\nabla^2f(x)_{j, j} + \nabla^2f(x)_{k, k} - \nabla^2f(x)_{j, k} - \nabla^2f(x)_{k, j}}\right\}.
\end{equation}
Lemma \ref{lemma:vem-qp} then leads to the following pseudocode of the vertex exchange method of solving the QP problem \eqref{eq:qp}, as presented in Algorithm \ref{alg:vem-qp}.

\begin{algorithm}[htb!]
\caption{The vertex exchange method for \eqref{eq:qp} \cite{liang2024vertex}.}\label{alg:vem-qp}
\begin{algorithmic}[1]
\State \textbf{Input:} Two feasible point $x,z\in\cF$ with $\nabla f(x)$ and $\nabla^2f(x)$, and a termination tolerance $\epsilon>0$.
\State Compute $\nabla q(z;x) = \nabla f(x) + \nabla^2f(x)(z-x)$.
\While{$z$ is not an $\epsilon$-solution of \eqref{eq:qp}} 
    \State Find the indices $j$ and $k$ by \eqref{eq:vemqpjk}.
    \State Compute the step size $\eta$ by \eqref{eq:etaqp}.
    \State Update $z$ by \eqref{eq:qp-newz}.
    \State Update the gradient $\nabla q(z;x) \leftarrow \nabla q(z;x) + \eta \left(\nabla^2f(x)_{:, j} - \nabla^2f(x)_{:,k}\right)$.
\EndWhile
\State \textbf{Output:} $x$.
\end{algorithmic}
\end{algorithm}

From the description of Algorithm \ref{alg:vem-qp}, we see that it admits extremely simple implementation and updating the gradient $\nabla q(\cdot;x)$ can be performed much more efficiently than updating the gradient $\nabla f(\cdot)$. More importantly, Algorithm \ref{alg:vem-qp} is globally convergent, as stated in the following theorem. 

\begin{theorem}[{\cite[Theorem 5]{liang2024vertex}}]
    \label{thm:vemqp}
    Suppose without loss of generality that Algorithm \ref{alg:vem-qp} generates an infinite sequence (otherwise, the algorithm produces an optimal solution in a finite number of iterations), then the whole sequence converges to the global optimal solution of problem \eqref{eq:qp}.
\end{theorem}

Overall speaking, we have leveraged the projected Newton method with the vertex exchange method and proposed a convergent algorithmic framework for efficiently solving the subproblems in the BnB method to compute the exact optimal experimental design. 

\section{Results} \label{sec:exp}

In this section, we shall conduct a set of numerical experiments on computing exact A-optimal and D-optimal experimental designs to validate the practical efficiency of the proposed framework, namely \texttt{PNOD}. The primary goals of our experimental study are mainly two-fold:
\begin{itemize}
    \item To demonstrate superior efficiency of combining the PN method and the vertex exchange method in evaluating the subproblems when compared to the direct application of the vertex exchange method alone, i.e., the \texttt{Co-BnB} method as implemented in \cite{hendrych2023solving}.
    \item To compare the ability of finding the exact optimal design subject to a time limit with the Frank-Wolfe-based mixed-integer convex method proposed in \cite{hendrych2023solving}, namely \texttt{Boscia}, which presents as one of the state-of-the-art methods in the literature. 
\end{itemize}
To this end, we employ the same experimental setup as in \cite{hendrych2023solving}. Specifically, we generate both independent and correlated data using random seeds from 1 to 5, representing by the matrix $A\in \mathbb{R}^{m\times n}$ with $m\in \{50, 60, 80, 100, 120\}$ and $n = m/10$. The experimental budget $N$ is set as $\left\lfloor 3n/2\right\rfloor $. We apply the greedy heuristic approach as presented in Algorithm \ref{alg:round} for generating a feasible solution for all the three methods, where the entries of the upper bound vector $\hat{u}\in\Z_+^m$ are randomly sampled from 1 to $N/3$. The time limit is 2 hours and the termination tolerances are set as $\texttt{abstol} = 10^{-2}$ and $\texttt{reltol} = 10^{-6}$, respectively.

Our numerical results are obtained from a machine with Ubuntu 22.04,  Intel i9-9820@3.3Ghz cores and 125GB of RAM. The detailed computational results can be found in Appendix \ref{appendix-exp}.  Our code is hosted at: 
\begin{center}
    \url{https://github.com/liangling98/PNOD.jl}
\end{center}

\subsection{Efficiency in evaluating nodes}\label{subsec-exp-nodes}
We first compare the performance in terms of the efficiency in evaluating nodes in the BnB framework. To this end, we record the number of nodes can be evaluated by each solver within one second. In particular, we divided the whole testing instances into four groups, each of which has the same criterion (A or D criterion) and the same type of data (independent or correlated data). For each $m$, we then compute the average number of nodes per second for each solver over the 5 seeds. 

\begin{figure}[htb!]
    \centering
    \includegraphics[width=0.45\linewidth]{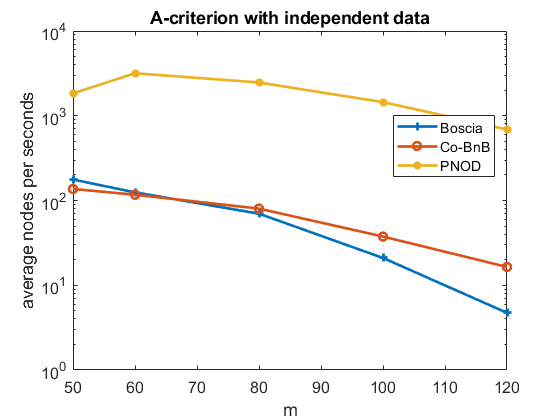}
    \includegraphics[width=0.45\linewidth]{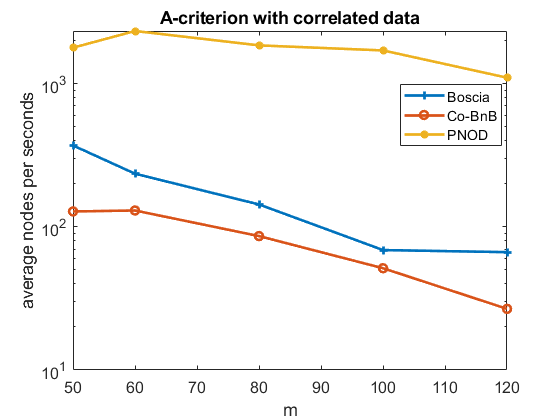}
    \includegraphics[width=0.45\linewidth]{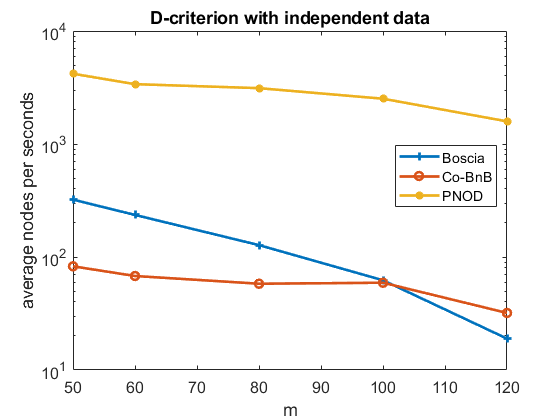}
    \includegraphics[width=0.45\linewidth]{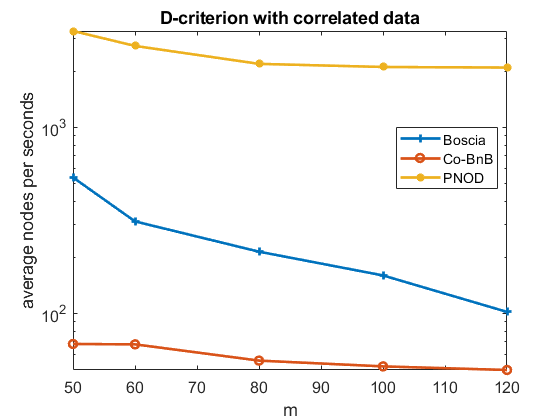}
    \caption{Average number of nodes per second (over 5 seeds) that processed by each solver.}
    \label{fig:nodespersecond}
\end{figure}

The computational results are shown in Figure \ref{fig:nodespersecond}, where it is evident that \texttt{PNOD} significantly outperforms both \texttt{Bosica} and \texttt{Co-BnB}. Specifically, \texttt{PNOD} is able to evaluate a substantially higher number of nodes within the BnB framework compared to the other two methods. This demonstrates that the use of the projected Newton framework in Algorithm \ref{alg:pn}, combined with the vertex exchange method from Algorithm \ref{alg:vem-qp}, offers a much more efficient approach for node evaluation than applying the vertex exchange method directly from Algorithm \ref{alg:vem-direct}. Additionally, it can be observed that for problems with independent data, \texttt{Bosica} and \texttt{Co-BnB} perform comparably. However, when dealing with correlated data, \texttt{Bosica} shows a clear advantage over \texttt{Co-BnB}.

\subsection{Efficiency in solving \eqref{eq:oedp}}\label{subsec-cpu}

Next, we shall compare the efficiency in solving the mixed-integer programming problem \eqref{eq:oedp} in terms of the total computational time taken by each solver subject to the same termination conditions as specified at the end of Section \ref{sec:bnb}. As already mentioned, we divided the whole testing instances into four groups, each of which has the same criterion and the same type of data. For each $m$, we compute the average computational time taken by each solver over the 5 seeds.

\begin{figure}[htb!]
    \centering
    \includegraphics[width=0.45\linewidth]{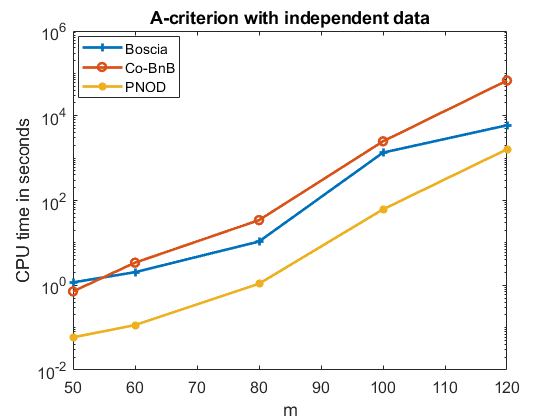}
    \includegraphics[width=0.45\linewidth]{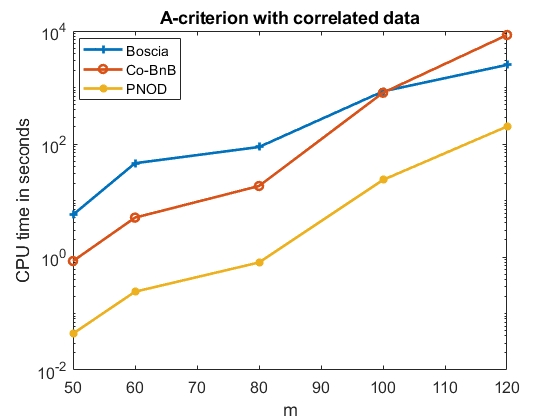}
    \includegraphics[width=0.45\linewidth]{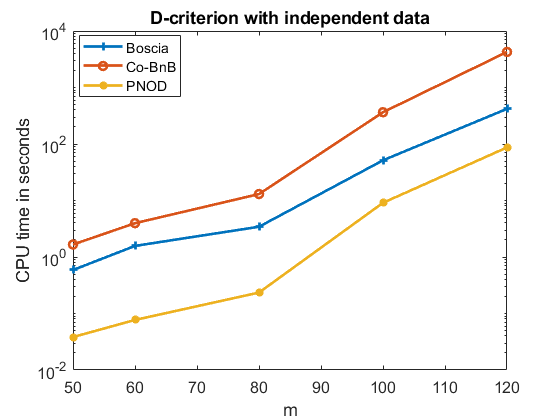}
    \includegraphics[width=0.45\linewidth]{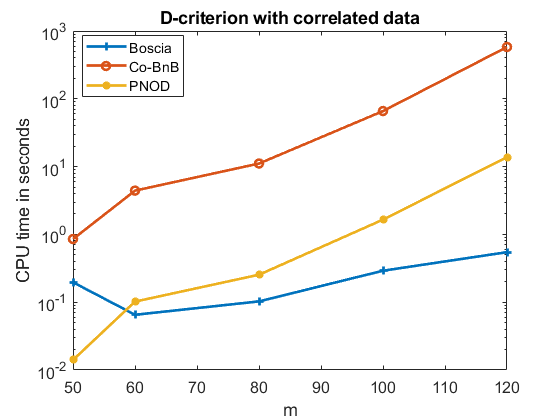}
    \caption{Average CPU time (over 5 seeds)  taken by each solver.}
    \label{fig:cpu}
\end{figure}

Figure \ref{fig:cpu} presents the average computational times, highlighting that \texttt{PNOD} outperforms \texttt{Co-BnB}, primarily due to the efficiency gained in solving the continuous optimization problem at each node within the BnB tree, as discussed in the previous subsection. Additionally, \texttt{PNOD} surpasses \texttt{Bosica} for A-optimal design with both independent and correlated data, as well as for D-optimal design with independent data. However, for D-optimal design with correlated data, \texttt{Bosica} outperforms \texttt{PNOD}.

The computational results also indicate that the A-optimal design with independent data is the most challenging problem class, as none of \texttt{Boscia} and \texttt{Co-BnB} were able to solve it within the two-hour time limit for $m=120$. Conversely, \texttt{PNOD} is able to successfully terminate within the time limit. Notably, \texttt{Co-BnB} fails to terminate correctly when the time limit is reached, which highlights a bug that needs addressing. In contrast, the D-optimal design with correlated data proves to be the easiest, with all three solvers terminating successfully within a short time frame.

\begin{figure}[htb!]
    \centering
    \includegraphics[width=0.75\linewidth]{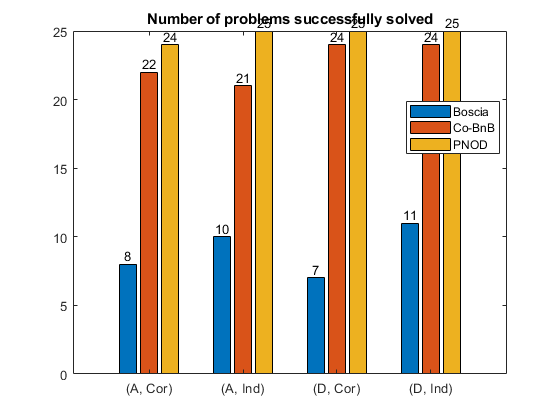}
    \caption{Comparison of solution quality in terms of objective values.}
    \label{fig:solved}
\end{figure}

However, computational time alone does not provide a complete picture of a solver's effectiveness in addressing \eqref{eq:oedp}. It is equally important to evaluate the solution quality produced by each solver. Figure \ref{fig:solved} illustrates the number of test instances 'successfully solved' by three solvers. Here, a problem is considered 'successfully solved' if the solver terminates before the time limit and returns a solution with a lower objective value.

From the results in Figure \ref{fig:solved}, we observe that while \texttt{Bosica} exhibits competitive performance in terms of total computational time, the solution quality it produces is generally inferior to that of \texttt{Co-BnB} and  \texttt{PNOD}. This, combined with its efficiency in computational time, underscores \texttt{PNOD}'s ability to not only solve problems faster but also deliver solutions with higher accuracy.

\section{Conclusions} \label{sec:conclusions}

We have enhanced the branch and bound (BnB) method for computing exact optimal experimental designs by incorporating the projected Newton method alongside the vertex exchange method. Instead of relying solely on the vertex exchange method for solving continuous relaxations in the search tree—an approach with unclear theoretical convergence guarantees—we have developed a novel framework that ensures robust convergence. This framework leverages the projected Newton method, with subproblems efficiently solved using the vertex exchange method. Extensive numerical experiments have demonstrated that our proposed framework outperforms state-of-the-art solvers in terms of efficiency and accuracy. Additionally, we have made available an open source Julia package, \texttt{PNOD.jl}, to facilitate potential practical applications and further research.

\section*{Acknowledgements}
The authors were partially supported by the US National Science Foundation under awards DMS-2244988, DMS-2206333, the Office of Naval Research Award N00014-23-1-2007, and the DARPA D24AP00325-00.
\clearpage

\appendix
\section{Detailed computational results}\label{appendix-exp}

\begin{longtable}{llllllllll}\\
\caption{Detailed computational results. \label{table-results}}
\\ \hline 
		Cri          & Cor        & m                   & n                  & N                  & seed               & Solver & Obj & Cpu & Nodes \\ \hline 
		\endfirsthead
		\multicolumn{10}{c}%
		{{\bfseries Table \thetable\ continued from previous page}} \\ \hline 
		Cri          & Cor        & m                   & n                  & N                  & seed               & Solver & Obj & Cpu & Nodes \\
        \hline 
		\endhead

 \multirow{3}{*}{A} & \multirow{3}{*}{0} & \multirow{3}{*}{50} & \multirow{3}{*}{5} & \multirow{3}{*}{7} & \multirow{3}{*}{1} & \texttt{Boscia} & 1.43476213e+00 &  3.050 & 81 \\
 & & & & & & \texttt{Co-BnB} & 1.43476213e+00 &  0.337 & 61 \\
 & & & & & & \texttt{PNOD} & 1.43476213e+00 &  0.067 & 57 \\ \hline
 \multirow{3}{*}{A} & \multirow{3}{*}{0} & \multirow{3}{*}{50} & \multirow{3}{*}{5} & \multirow{3}{*}{7} & \multirow{3}{*}{2} & \texttt{Boscia} & 1.44263033e+00 &  0.092 & 21 \\
 & & & & & & \texttt{Co-BnB} & 1.44263033e+00 &  0.327 & 25 \\
 & & & & & & \texttt{PNOD} & 1.44263033e+00 &  0.014 & 25 \\ \hline
 \multirow{3}{*}{A} & \multirow{3}{*}{0} & \multirow{3}{*}{50} & \multirow{3}{*}{5} & \multirow{3}{*}{7} & \multirow{3}{*}{3} & \texttt{Boscia} & 1.59991121e+00 &  1.427 & 325 \\
 & & & & & & \texttt{Co-BnB} & 1.59991121e+00 &  1.650 & 221 \\
 & & & & & & \texttt{PNOD} & 1.59991121e+00 &  0.132 & 245 \\ \hline
 \multirow{3}{*}{A} & \multirow{3}{*}{0} & \multirow{3}{*}{50} & \multirow{3}{*}{5} & \multirow{3}{*}{7} & \multirow{3}{*}{4} & \texttt{Boscia} & 1.59861820e+00 &  1.104 & 227 \\
 & & & & & & \texttt{Co-BnB} & 1.57682686e+00 &  1.049 & 153 \\
 & & & & & & \texttt{PNOD} & 1.57682686e+00 &  0.059 & 195 \\ \hline
 \multirow{3}{*}{A} & \multirow{3}{*}{0} & \multirow{3}{*}{50} & \multirow{3}{*}{5} & \multirow{3}{*}{7} & \multirow{3}{*}{5} & \texttt{Boscia} & 1.56859177e+00 &  0.169 & 33 \\
 & & & & & & \texttt{Co-BnB} & 1.56859177e+00 &  0.212 & 31 \\
 & & & & & & \texttt{PNOD} & 1.56859177e+00 &  0.023 & 31 \\ \hline
 \multirow{3}{*}{A} & \multirow{3}{*}{0} & \multirow{3}{*}{60} & \multirow{3}{*}{6} & \multirow{3}{*}{9} & \multirow{3}{*}{1} & \texttt{Boscia} & 1.60523689e+00 &  2.348 & 303 \\
 & & & & & & \texttt{Co-BnB} & 1.60523689e+00 &  2.231 & 267 \\
 & & & & & & \texttt{PNOD} & 1.60523689e+00 &  0.083 & 247 \\ \hline
 \multirow{3}{*}{A} & \multirow{3}{*}{0} & \multirow{3}{*}{60} & \multirow{3}{*}{6} & \multirow{3}{*}{9} & \multirow{3}{*}{2} & \texttt{Boscia} & 1.66300992e+00 &  2.654 & 199 \\
 & & & & & & \texttt{Co-BnB} & 1.60573571e+00 &  2.329 & 273 \\
 & & & & & & \texttt{PNOD} & 1.60573571e+00 &  0.085 & 281 \\ \hline
 \multirow{3}{*}{A} & \multirow{3}{*}{0} & \multirow{3}{*}{60} & \multirow{3}{*}{6} & \multirow{3}{*}{9} & \multirow{3}{*}{3} & \texttt{Boscia} & 1.64878270e+00 &  1.716 & 255 \\
 & & & & & & \texttt{Co-BnB} & 1.64878270e+00 &  1.846 & 235 \\
 & & & & & & \texttt{PNOD} & 1.64878270e+00 &  0.077 & 223 \\ \hline
 \multirow{3}{*}{A} & \multirow{3}{*}{0} & \multirow{3}{*}{60} & \multirow{3}{*}{6} & \multirow{3}{*}{9} & \multirow{3}{*}{4} & \texttt{Boscia} & 1.70293533e+00 &  2.180 & 259 \\
 & & & & & & \texttt{Co-BnB} & 1.68441098e+00 &  6.097 & 513 \\
 & & & & & & \texttt{PNOD} & 1.68441098e+00 &  0.185 & 537 \\ \hline
 \multirow{3}{*}{A} & \multirow{3}{*}{0} & \multirow{3}{*}{60} & \multirow{3}{*}{6} & \multirow{3}{*}{9} & \multirow{3}{*}{5} & \texttt{Boscia} & 1.62278263e+00 &  1.283 & 195 \\
 & & & & & & \texttt{Co-BnB} & 1.62278263e+00 &  4.556 & 621 \\
 & & & & & & \texttt{PNOD} & 1.62278263e+00 &  0.144 & 545 \\ \hline
 \multirow{3}{*}{A} & \multirow{3}{*}{0} & \multirow{3}{*}{80} & \multirow{3}{*}{8} & \multirow{3}{*}{12} & \multirow{3}{*}{1} & \texttt{Boscia} & 1.65705252e+00 &  8.468 & 615 \\
 & & & & & & \texttt{Co-BnB} & 1.64545515e+00 & 26.013 & 2207 \\
 & & & & & & \texttt{PNOD} & 1.64545515e+00 &  0.852 & 2239 \\ \hline
 \multirow{3}{*}{A} & \multirow{3}{*}{0} & \multirow{3}{*}{80} & \multirow{3}{*}{8} & \multirow{3}{*}{12} & \multirow{3}{*}{2} & \texttt{Boscia} & 1.84222385e+00 & 14.997 & 717 \\
 & & & & & & \texttt{Co-BnB} & 1.80714687e+00 & 18.261 & 1433 \\
 & & & & & & \texttt{PNOD} & 1.80714687e+00 &  0.600 & 1321 \\ \hline
 \multirow{3}{*}{A} & \multirow{3}{*}{0} & \multirow{3}{*}{80} & \multirow{3}{*}{8} & \multirow{3}{*}{12} & \multirow{3}{*}{3} & \texttt{Boscia} & 1.78026518e+00 &  6.610 & 469 \\
 & & & & & & \texttt{Co-BnB} & 1.78026518e+00 & 64.576 & 5009 \\
 & & & & & & \texttt{PNOD} & 1.78026518e+00 &  1.933 & 4965 \\ \hline
 \multirow{3}{*}{A} & \multirow{3}{*}{0} & \multirow{3}{*}{80} & \multirow{3}{*}{8} & \multirow{3}{*}{12} & \multirow{3}{*}{4} & \texttt{Boscia} & 1.75371602e+00 &  8.997 & 695 \\
 & & & & & & \texttt{Co-BnB} & 1.75104278e+00 & 33.144 & 2981 \\
 & & & & & & \texttt{PNOD} & 1.75104278e+00 &  1.049 & 3029 \\ \hline
 \multirow{3}{*}{A} & \multirow{3}{*}{0} & \multirow{3}{*}{80} & \multirow{3}{*}{8} & \multirow{3}{*}{12} & \multirow{3}{*}{5} & \texttt{Boscia} & 1.76460705e+00 & 14.782 & 1193 \\
 & & & & & & \texttt{Co-BnB} & 1.76460705e+00 & 30.120 & 2109 \\
 & & & & & & \texttt{PNOD} & 1.76460705e+00 &  0.999 & 2053 \\ \hline
 \multirow{3}{*}{A} & \multirow{3}{*}{0} & \multirow{3}{*}{100} & \multirow{3}{*}{10} & \multirow{3}{*}{15} & \multirow{3}{*}{1} & \texttt{Boscia} & 1.86193106e+00 & 48.052 & 1401 \\
 & & & & & & \texttt{Co-BnB} & 1.86193106e+00 & 1069.280 & 46529 \\
 & & & & & & \texttt{PNOD} & 1.86193106e+00 & 27.872 & 46069 \\ \hline
 \multirow{3}{*}{A} & \multirow{3}{*}{0} & \multirow{3}{*}{100} & \multirow{3}{*}{10} & \multirow{3}{*}{15} & \multirow{3}{*}{2} & \texttt{Boscia} & 1.93240637e+00 & 45.612 & 1195 \\
 & & & & & & \texttt{Co-BnB} & 1.92995032e+00 & 620.547 & 22897 \\
 & & & & & & \texttt{PNOD} & 1.92995032e+00 & 15.655 & 23011 \\ \hline
 \multirow{3}{*}{A} & \multirow{3}{*}{0} & \multirow{3}{*}{100} & \multirow{3}{*}{10} & \multirow{3}{*}{15} & \multirow{3}{*}{3} & \texttt{Boscia} & 1.93920423e+00 & 3863.056 & 55733 \\
 & & & & & & \texttt{Co-BnB} & 1.90124258e+00 & 3012.128 & 113965 \\
 & & & & & & \texttt{PNOD} & 1.90124258e+00 & 76.937 & 104773 \\ \hline
 \multirow{3}{*}{A} & \multirow{3}{*}{0} & \multirow{3}{*}{100} & \multirow{3}{*}{10} & \multirow{3}{*}{15} & \multirow{3}{*}{4} & \texttt{Boscia} & 1.91286729e+00 & 2326.913 & 28649 \\
 & & & & & & \texttt{Co-BnB} & 1.87387011e+00 & 6138.169 & 205181 \\
 & & & & & & \texttt{PNOD} & 1.87387011e+00 & 145.031 & 205277 \\ \hline
 \multirow{3}{*}{A} & \multirow{3}{*}{0} & \multirow{3}{*}{100} & \multirow{3}{*}{10} & \multirow{3}{*}{15} & \multirow{3}{*}{5} & \texttt{Boscia} & 1.87353603e+00 & 466.443 & 10631 \\
 & & & & & & \texttt{Co-BnB} & 1.86486311e+00 & 1616.935 & 58137 \\
 & & & & & & \texttt{PNOD} & 1.86486311e+00 & 43.882 & 58557 \\ \hline
 \multirow{3}{*}{A} & \multirow{3}{*}{0} & \multirow{3}{*}{120} & \multirow{3}{*}{12} & \multirow{3}{*}{18} & \multirow{3}{*}{1} & \texttt{Boscia} & 1.94702588e+00 & 5619.762 & 20077 \\
 & & & & & & \texttt{Co-BnB} & 1.90899799e+00 & 49249.802 & 940673 \\
 & & & & & & \texttt{PNOD} & 1.90899799e+00 & 1083.183 & 942827 \\ \hline
 \multirow{3}{*}{A} & \multirow{3}{*}{0} & \multirow{3}{*}{120} & \multirow{3}{*}{12} & \multirow{3}{*}{18} & \multirow{3}{*}{2} & \texttt{Boscia} & 2.03734591e+00 & 5011.477 & 19747 \\
 & & & & & & \texttt{Co-BnB} & 1.98338464e+00 & 75262.808 & 1294837 \\
 & & & & & & \texttt{PNOD} & 1.98338464e+00 & 1768.700 & 1282173 \\ \hline
 \multirow{3}{*}{A} & \multirow{3}{*}{0} & \multirow{3}{*}{120} & \multirow{3}{*}{12} & \multirow{3}{*}{18} & \multirow{3}{*}{3} & \texttt{Boscia} & 1.95304628e+00 & 7200.015 & 41301 \\
 & & & & & & \texttt{Co-BnB} & 1.91819177e+00 & 100666.256 & 1317025 \\
 & & & & & & \texttt{PNOD} & 1.91819177e+00 & 2502.419 & 1313579 \\ \hline
 \multirow{3}{*}{A} & \multirow{3}{*}{0} & \multirow{3}{*}{120} & \multirow{3}{*}{12} & \multirow{3}{*}{18} & \multirow{3}{*}{4} & \texttt{Boscia} & 2.01711819e+00 & 7200.014 & 40249 \\
 & & & & & & \texttt{Co-BnB} & 1.99251646e+00 & 108573.788 & 1259741 \\
 & & & & & & \texttt{PNOD} & 1.99251646e+00 & 2615.282 & 1260691 \\ \hline
 \multirow{3}{*}{A} & \multirow{3}{*}{0} & \multirow{3}{*}{120} & \multirow{3}{*}{12} & \multirow{3}{*}{18} & \multirow{3}{*}{5} & \texttt{Boscia} & 1.92720605e+00 & 4764.300 & 22767 \\
 & & & & & & \texttt{Co-BnB} & 1.87266188e+00 & 4932.900 & 104443 \\
 & & & & & & \texttt{PNOD} & 1.87266188e+00 & 123.760 & 103843 \\ \hline
 \multirow{3}{*}{A} & \multirow{3}{*}{1} & \multirow{3}{*}{50} & \multirow{3}{*}{5} & \multirow{3}{*}{7} & \multirow{3}{*}{1} & \texttt{Boscia} & -2.36392004e+00 &  8.831 & 2811 \\
 & & & & & & \texttt{Co-BnB} & -2.36392004e+00 &  2.312 & 93 \\
 & & & & & & \texttt{PNOD} & -2.36392004e+00 &  0.027 & 77 \\ \hline
 \multirow{3}{*}{A} & \multirow{3}{*}{1} & \multirow{3}{*}{50} & \multirow{3}{*}{5} & \multirow{3}{*}{7} & \multirow{3}{*}{2} & \texttt{Boscia} & -2.37420613e+00 &  2.931 & 1081 \\
 & & & & & & \texttt{Co-BnB} & -2.37420613e+00 &  0.366 & 67 \\
 & & & & & & \texttt{PNOD} & -2.37420613e+00 &  0.043 & 71 \\ \hline
 \multirow{3}{*}{A} & \multirow{3}{*}{1} & \multirow{3}{*}{50} & \multirow{3}{*}{5} & \multirow{3}{*}{7} & \multirow{3}{*}{3} & \texttt{Boscia} & -2.41029371e+00 &  5.854 & 2193 \\
 & & & & & & \texttt{Co-BnB} & -2.41029371e+00 &  0.461 & 61 \\
 & & & & & & \texttt{PNOD} & -2.41029371e+00 &  0.040 & 55 \\ \hline
 \multirow{3}{*}{A} & \multirow{3}{*}{1} & \multirow{3}{*}{50} & \multirow{3}{*}{5} & \multirow{3}{*}{7} & \multirow{3}{*}{4} & \texttt{Boscia} & -2.51503197e+00 &  3.856 & 1321 \\
 & & & & & & \texttt{Co-BnB} & -2.51503197e+00 &  0.638 & 81 \\
 & & & & & & \texttt{PNOD} & -2.48532997e+00 &  0.051 & 89 \\ \hline
 \multirow{3}{*}{A} & \multirow{3}{*}{1} & \multirow{3}{*}{50} & \multirow{3}{*}{5} & \multirow{3}{*}{7} & \multirow{3}{*}{5} & \texttt{Boscia} & -2.73926296e+00 &  6.524 & 2927 \\
 & & & & & & \texttt{Co-BnB} & -2.77613169e+00 &  0.461 & 73 \\
 & & & & & & \texttt{PNOD} & -2.77613169e+00 &  0.060 & 77 \\ \hline
 \multirow{3}{*}{A} & \multirow{3}{*}{1} & \multirow{3}{*}{60} & \multirow{3}{*}{6} & \multirow{3}{*}{9} & \multirow{3}{*}{1} & \texttt{Boscia} & -2.41066388e+00 & 22.948 & 5425 \\
 & & & & & & \texttt{Co-BnB} & -2.41066388e+00 &  3.625 & 531 \\
 & & & & & & \texttt{PNOD} & -2.45069804e+00 &  0.024 & 61 \\ \hline
 \multirow{3}{*}{A} & \multirow{3}{*}{1} & \multirow{3}{*}{60} & \multirow{3}{*}{6} & \multirow{3}{*}{9} & \multirow{3}{*}{2} & \texttt{Boscia} & -2.55835815e+00 & 65.580 & 14431 \\
 & & & & & & \texttt{Co-BnB} & -2.55835815e+00 &  8.812 & 1153 \\
 & & & & & & \texttt{PNOD} & -2.55835815e+00 &  0.482 & 1135 \\ \hline
 \multirow{3}{*}{A} & \multirow{3}{*}{1} & \multirow{3}{*}{60} & \multirow{3}{*}{6} & \multirow{3}{*}{9} & \multirow{3}{*}{3} & \texttt{Boscia} & -2.60824033e+00 & 10.133 & 2609 \\
 & & & & & & \texttt{Co-BnB} & -2.66835809e+00 &  1.152 & 97 \\
 & & & & & & \texttt{PNOD} & -2.66835809e+00 &  0.054 & 101 \\ \hline
 \multirow{3}{*}{A} & \multirow{3}{*}{1} & \multirow{3}{*}{60} & \multirow{3}{*}{6} & \multirow{3}{*}{9} & \multirow{3}{*}{4} & \texttt{Boscia} & -2.64358105e+00 & 80.491 & 16851 \\
 & & & & & & \texttt{Co-BnB} & -2.64358105e+00 &  6.916 & 897 \\
 & & & & & & \texttt{PNOD} & -2.64358105e+00 &  0.337 & 843 \\ \hline
 \multirow{3}{*}{A} & \multirow{3}{*}{1} & \multirow{3}{*}{60} & \multirow{3}{*}{6} & \multirow{3}{*}{9} & \multirow{3}{*}{5} & \texttt{Boscia} & -2.65236178e+00 & 49.392 & 12501 \\
 & & & & & & \texttt{Co-BnB} & -2.65236178e+00 &  4.366 & 697 \\
 & & & & & & \texttt{PNOD} & -2.65236178e+00 &  0.318 & 765 \\ \hline
 \multirow{3}{*}{A} & \multirow{3}{*}{1} & \multirow{3}{*}{80} & \multirow{3}{*}{8} & \multirow{3}{*}{12} & \multirow{3}{*}{1} & \texttt{Boscia} & -2.69676476e+00 & 50.390 & 7807 \\
 & & & & & & \texttt{Co-BnB} & -2.77985123e+00 & 10.497 & 1185 \\
 & & & & & & \texttt{PNOD} & -2.77985123e+00 &  0.588 & 1337 \\ \hline
 \multirow{3}{*}{A} & \multirow{3}{*}{1} & \multirow{3}{*}{80} & \multirow{3}{*}{8} & \multirow{3}{*}{12} & \multirow{3}{*}{2} & \texttt{Boscia} & -2.36783404e+00 & 80.228 & 13871 \\
 & & & & & & \texttt{Co-BnB} & -2.36783404e+00 & 34.301 & 3125 \\
 & & & & & & \texttt{PNOD} & -2.36783404e+00 &  1.392 & 3161 \\ \hline
 \multirow{3}{*}{A} & \multirow{3}{*}{1} & \multirow{3}{*}{80} & \multirow{3}{*}{8} & \multirow{3}{*}{12} & \multirow{3}{*}{3} & \texttt{Boscia} & -2.37587377e+00 & 132.261 & 16875 \\
 & & & & & & \texttt{Co-BnB} & -2.46304846e+00 & 17.067 & 1193 \\
 & & & & & & \texttt{PNOD} & -2.46304846e+00 &  0.663 & 1175 \\ \hline
 \multirow{3}{*}{A} & \multirow{3}{*}{1} & \multirow{3}{*}{80} & \multirow{3}{*}{8} & \multirow{3}{*}{12} & \multirow{3}{*}{4} & \texttt{Boscia} & -2.57955302e+00 & 12.467 & 1643 \\
 & & & & & & \texttt{Co-BnB} & -2.88951621e+00 &  3.164 & 235 \\
 & & & & & & \texttt{PNOD} & -2.88951621e+00 &  0.381 & 235 \\ \hline
 \multirow{3}{*}{A} & \multirow{3}{*}{1} & \multirow{3}{*}{80} & \multirow{3}{*}{8} & \multirow{3}{*}{12} & \multirow{3}{*}{5} & \texttt{Boscia} & -2.59731739e+00 & 167.303 & 21779 \\
 & & & & & & \texttt{Co-BnB} & -2.63299439e+00 & 25.280 & 2099 \\
 & & & & & & \texttt{PNOD} & -2.63299439e+00 &  0.978 & 2289 \\ \hline
 \multirow{3}{*}{A} & \multirow{3}{*}{1} & \multirow{3}{*}{100} & \multirow{3}{*}{10} & \multirow{3}{*}{15} & \multirow{3}{*}{1} & \texttt{Boscia} & -2.69328018e+00 & 108.694 & 7109 \\
 & & & & & & \texttt{Co-BnB} & -2.82601776e+00 & 596.173 & 33173 \\
 & & & & & & \texttt{PNOD} & -2.82601776e+00 & 18.561 & 33005 \\ \hline
 \multirow{3}{*}{A} & \multirow{3}{*}{1} & \multirow{3}{*}{100} & \multirow{3}{*}{10} & \multirow{3}{*}{15} & \multirow{3}{*}{2} & \texttt{Boscia} & -2.60099564e+00 & 2198.524 & 108199 \\
 & & & & & & \texttt{Co-BnB} & -2.61868441e+00 & 1209.141 & 53481 \\
 & & & & & & \texttt{PNOD} & -2.61868441e+00 & 33.033 & 51969 \\ \hline
 \multirow{3}{*}{A} & \multirow{3}{*}{1} & \multirow{3}{*}{100} & \multirow{3}{*}{10} & \multirow{3}{*}{15} & \multirow{3}{*}{3} & \texttt{Boscia} & -2.73059530e+00 & 294.151 & 25961 \\
 & & & & & & \texttt{Co-BnB} & -2.76678944e+00 & 192.389 & 11015 \\
 & & & & & & \texttt{PNOD} & -2.76678944e+00 &  4.597 & 8245 \\ \hline
 \multirow{3}{*}{A} & \multirow{3}{*}{1} & \multirow{3}{*}{100} & \multirow{3}{*}{10} & \multirow{3}{*}{15} & \multirow{3}{*}{4} & \texttt{Boscia} & -2.91438132e+00 & 352.687 & 28291 \\
 & & & & & & \texttt{Co-BnB} & -2.93653025e+00 & 1299.707 & 63833 \\
 & & & & & & \texttt{PNOD} & -2.93653025e+00 & 37.264 & 64085 \\ \hline
 \multirow{3}{*}{A} & \multirow{3}{*}{1} & \multirow{3}{*}{100} & \multirow{3}{*}{10} & \multirow{3}{*}{15} & \multirow{3}{*}{5} & \texttt{Boscia} & -2.68003361e+00 & 1318.424 & 80909 \\
 & & & & & & \texttt{Co-BnB} & -2.71881090e+00 & 741.783 & 37729 \\
 & & & & & & \texttt{PNOD} & -2.71881090e+00 & 22.330 & 37633 \\ \hline
 \multirow{3}{*}{A} & \multirow{3}{*}{1} & \multirow{3}{*}{120} & \multirow{3}{*}{12} & \multirow{3}{*}{18} & \multirow{3}{*}{1} & \texttt{Boscia} & -2.93726125e+00 & 2240.748 & 76641 \\
 & & & & & & \texttt{Co-BnB} & -2.95758584e+00 & 2609.847 & 89805 \\
 & & & & & & \texttt{PNOD} & -2.95758584e+00 & 59.335 & 86529 \\ \hline
 \multirow{3}{*}{A} & \multirow{3}{*}{1} & \multirow{3}{*}{120} & \multirow{3}{*}{12} & \multirow{3}{*}{18} & \multirow{3}{*}{2} & \texttt{Boscia} & -2.74685173e+00 & 2139.216 & 61661 \\
 & & & & & & \texttt{Co-BnB} & -2.80626769e+00 & 11783.900 & 297535 \\
 & & & & & & \texttt{PNOD} & -2.80626769e+00 & 258.825 & 263489 \\ \hline
 \multirow{3}{*}{A} & \multirow{3}{*}{1} & \multirow{3}{*}{120} & \multirow{3}{*}{12} & \multirow{3}{*}{18} & \multirow{3}{*}{3} & \texttt{Boscia} & -2.75848240e+00 & 6508.095 & 100013 \\
 & & & & & & \texttt{Co-BnB} & -2.89197885e+00 & 19067.738 & 501067 \\
 & & & & & & \texttt{PNOD} & -2.89197885e+00 & 467.973 & 525355 \\ \hline
 \multirow{3}{*}{A} & \multirow{3}{*}{1} & \multirow{3}{*}{120} & \multirow{3}{*}{12} & \multirow{3}{*}{18} & \multirow{3}{*}{4} & \texttt{Boscia} & -2.48922663e+00 &  0.169 & 41 \\
 & & & & & & \texttt{Co-BnB} & -3.00432476e+00 & 5299.392 & 124181 \\
 & & & & & & \texttt{PNOD} & -3.00432476e+00 & 145.200 & 133445 \\ \hline
 \multirow{3}{*}{A} & \multirow{3}{*}{1} & \multirow{3}{*}{120} & \multirow{3}{*}{12} & \multirow{3}{*}{18} & \multirow{3}{*}{5} & \texttt{Boscia} & -2.74659652e+00 & 1676.871 & 19981 \\
 & & & & & & \texttt{Co-BnB} & -2.93910296e+00 & 3717.354 & 90543 \\
 & & & & & & \texttt{PNOD} & -2.93910296e+00 & 90.021 & 89753 \\ \hline
 \multirow{3}{*}{D} & \multirow{3}{*}{0} & \multirow{3}{*}{50} & \multirow{3}{*}{5} & \multirow{3}{*}{7} & \multirow{3}{*}{1} & \texttt{Boscia} & -4.16444318e-02 &  1.319 & 151 \\
 & & & & & & \texttt{Co-BnB} & -4.16444318e-02 &  1.059 & 101 \\
 & & & & & & \texttt{PNOD} & -4.16444320e-02 &  0.028 & 101 \\ \hline
 \multirow{3}{*}{D} & \multirow{3}{*}{0} & \multirow{3}{*}{50} & \multirow{3}{*}{5} & \multirow{3}{*}{7} & \multirow{3}{*}{2} & \texttt{Boscia} & -4.02420151e-02 &  0.381 & 143 \\
 & & & & & & \texttt{Co-BnB} & -4.02420151e-02 &  1.688 & 117 \\
 & & & & & & \texttt{PNOD} & -4.02420150e-02 &  0.016 & 95 \\ \hline
 \multirow{3}{*}{D} & \multirow{3}{*}{0} & \multirow{3}{*}{50} & \multirow{3}{*}{5} & \multirow{3}{*}{7} & \multirow{3}{*}{3} & \texttt{Boscia} & -3.46927855e-02 &  0.899 & 341 \\
 & & & & & & \texttt{Co-BnB} & -3.46927855e-02 &  3.020 & 237 \\
 & & & & & & \texttt{PNOD} & -3.46927860e-02 &  0.099 & 269 \\ \hline
 \multirow{3}{*}{D} & \multirow{3}{*}{0} & \multirow{3}{*}{50} & \multirow{3}{*}{5} & \multirow{3}{*}{7} & \multirow{3}{*}{4} & \texttt{Boscia} & -3.42516705e-02 &  0.192 & 81 \\
 & & & & & & \texttt{Co-BnB} & -3.45958189e-02 &  2.222 & 155 \\
 & & & & & & \texttt{PNOD} & -3.45958190e-02 &  0.041 & 163 \\ \hline
 \multirow{3}{*}{D} & \multirow{3}{*}{0} & \multirow{3}{*}{50} & \multirow{3}{*}{5} & \multirow{3}{*}{7} & \multirow{3}{*}{5} & \texttt{Boscia} & -3.69800870e-02 &  0.154 & 49 \\
 & & & & & & \texttt{Co-BnB} & -3.69800870e-02 &  0.328 & 33 \\
 & & & & & & \texttt{PNOD} & -3.69800870e-02 &  0.007 & 33 \\ \hline
 \multirow{3}{*}{D} & \multirow{3}{*}{0} & \multirow{3}{*}{60} & \multirow{3}{*}{6} & \multirow{3}{*}{9} & \multirow{3}{*}{1} & \texttt{Boscia} & -4.98856600e-02 &  2.447 & 511 \\
 & & & & & & \texttt{Co-BnB} & -4.98856600e-02 &  5.304 & 451 \\
 & & & & & & \texttt{PNOD} & -4.98856600e-02 &  0.121 & 439 \\ \hline
 \multirow{3}{*}{D} & \multirow{3}{*}{0} & \multirow{3}{*}{60} & \multirow{3}{*}{6} & \multirow{3}{*}{9} & \multirow{3}{*}{2} & \texttt{Boscia} & -5.10930680e-02 &  0.910 & 201 \\
 & & & & & & \texttt{Co-BnB} & -5.31752429e-02 &  2.555 & 131 \\
 & & & & & & \texttt{PNOD} & -5.31752430e-02 &  0.033 & 87 \\ \hline
 \multirow{3}{*}{D} & \multirow{3}{*}{0} & \multirow{3}{*}{60} & \multirow{3}{*}{6} & \multirow{3}{*}{9} & \multirow{3}{*}{3} & \texttt{Boscia} & -4.46260968e-02 &  1.555 & 377 \\
 & & & & & & \texttt{Co-BnB} & -4.46260968e-02 &  4.863 & 345 \\
 & & & & & & \texttt{PNOD} & -4.46260970e-02 &  0.118 & 365 \\ \hline
 \multirow{3}{*}{D} & \multirow{3}{*}{0} & \multirow{3}{*}{60} & \multirow{3}{*}{6} & \multirow{3}{*}{9} & \multirow{3}{*}{4} & \texttt{Boscia} & -4.35073925e-02 &  1.581 & 383 \\
 & & & & & & \texttt{Co-BnB} & -4.50922044e-02 &  1.813 & 139 \\
 & & & & & & \texttt{PNOD} & -4.50922040e-02 &  0.038 & 139 \\ \hline
 \multirow{3}{*}{D} & \multirow{3}{*}{0} & \multirow{3}{*}{60} & \multirow{3}{*}{6} & \multirow{3}{*}{9} & \multirow{3}{*}{5} & \texttt{Boscia} & -4.94525636e-02 &  1.369 & 361 \\
 & & & & & & \texttt{Co-BnB} & -4.94525636e-02 &  5.313 & 295 \\
 & & & & & & \texttt{PNOD} & -4.94525640e-02 &  0.075 & 291 \\ \hline
 \multirow{3}{*}{D} & \multirow{3}{*}{0} & \multirow{3}{*}{80} & \multirow{3}{*}{8} & \multirow{3}{*}{12} & \multirow{3}{*}{1} & \texttt{Boscia} & -7.09994854e-02 &  2.009 & 195 \\
 & & & & & & \texttt{Co-BnB} & -7.28698020e-02 & 10.543 & 639 \\
 & & & & & & \texttt{PNOD} & -7.28698020e-02 &  0.208 & 529 \\ \hline
 \multirow{3}{*}{D} & \multirow{3}{*}{0} & \multirow{3}{*}{80} & \multirow{3}{*}{8} & \multirow{3}{*}{12} & \multirow{3}{*}{2} & \texttt{Boscia} & -5.64598762e-02 &  4.854 & 455 \\
 & & & & & & \texttt{Co-BnB} & -5.73744886e-02 & 21.009 & 1197 \\
 & & & & & & \texttt{PNOD} & -5.73744890e-02 &  0.424 & 1179 \\ \hline
 \multirow{3}{*}{D} & \multirow{3}{*}{0} & \multirow{3}{*}{80} & \multirow{3}{*}{8} & \multirow{3}{*}{12} & \multirow{3}{*}{3} & \texttt{Boscia} & -6.34049118e-02 &  6.915 & 877 \\
 & & & & & & \texttt{Co-BnB} & -6.43260320e-02 & 23.619 & 1185 \\
 & & & & & & \texttt{PNOD} & -6.43260320e-02 &  0.352 & 1289 \\ \hline
 \multirow{3}{*}{D} & \multirow{3}{*}{0} & \multirow{3}{*}{80} & \multirow{3}{*}{8} & \multirow{3}{*}{12} & \multirow{3}{*}{4} & \texttt{Boscia} & -6.73685804e-02 &  1.690 & 255 \\
 & & & & & & \texttt{Co-BnB} & -6.79086308e-02 &  5.852 & 369 \\
 & & & & & & \texttt{PNOD} & -6.79086310e-02 &  0.104 & 373 \\ \hline
 \multirow{3}{*}{D} & \multirow{3}{*}{0} & \multirow{3}{*}{80} & \multirow{3}{*}{8} & \multirow{3}{*}{12} & \multirow{3}{*}{5} & \texttt{Boscia} & -6.41476149e-02 &  1.694 & 285 \\
 & & & & & & \texttt{Co-BnB} & -6.41476149e-02 &  4.035 & 239 \\
 & & & & & & \texttt{PNOD} & -6.41476150e-02 &  0.087 & 263 \\ \hline
 \multirow{3}{*}{D} & \multirow{3}{*}{0} & \multirow{3}{*}{100} & \multirow{3}{*}{10} & \multirow{3}{*}{15} & \multirow{3}{*}{1} & \texttt{Boscia} & -7.64891906e-02 & 22.680 & 1415 \\
 & & & & & & \texttt{Co-BnB} & -7.69236210e-02 & 263.522 & 14429 \\
 & & & & & & \texttt{PNOD} & -7.69236210e-02 &  5.677 & 14685 \\ \hline
 \multirow{3}{*}{D} & \multirow{3}{*}{0} & \multirow{3}{*}{100} & \multirow{3}{*}{10} & \multirow{3}{*}{15} & \multirow{3}{*}{2} & \texttt{Boscia} & -7.01315650e-02 & 10.920 & 725 \\
 & & & & & & \texttt{Co-BnB} & -7.07438243e-02 & 109.939 & 6115 \\
 & & & & & & \texttt{PNOD} & -7.07438240e-02 &  1.881 & 5643 \\ \hline
 \multirow{3}{*}{D} & \multirow{3}{*}{0} & \multirow{3}{*}{100} & \multirow{3}{*}{10} & \multirow{3}{*}{15} & \multirow{3}{*}{3} & \texttt{Boscia} & -7.13550481e-02 & 43.414 & 2597 \\
 & & & & & & \texttt{Co-BnB} & -7.13550481e-02 & 438.957 & 24535 \\
 & & & & & & \texttt{PNOD} & -7.13550480e-02 & 10.551 & 23587 \\ \hline
 \multirow{3}{*}{D} & \multirow{3}{*}{0} & \multirow{3}{*}{100} & \multirow{3}{*}{10} & \multirow{3}{*}{15} & \multirow{3}{*}{4} & \texttt{Boscia} & -7.39678375e-02 & 26.338 & 1573 \\
 & & & & & & \texttt{Co-BnB} & -7.39678375e-02 & 393.903 & 28735 \\
 & & & & & & \texttt{PNOD} & -7.39678370e-02 & 11.164 & 28707 \\ \hline
 \multirow{3}{*}{D} & \multirow{3}{*}{0} & \multirow{3}{*}{100} & \multirow{3}{*}{10} & \multirow{3}{*}{15} & \multirow{3}{*}{5} & \texttt{Boscia} & -7.41566590e-02 & 154.039 & 9659 \\
 & & & & & & \texttt{Co-BnB} & -7.41566590e-02 & 618.606 & 35089 \\
 & & & & & & \texttt{PNOD} & -7.41566590e-02 & 16.302 & 35581 \\ \hline
 \multirow{3}{*}{D} & \multirow{3}{*}{0} & \multirow{3}{*}{120} & \multirow{3}{*}{12} & \multirow{3}{*}{18} & \multirow{3}{*}{1} & \texttt{Boscia} & -8.51461833e-02 & 408.260 & 5795 \\
 & & & & & & \texttt{Co-BnB} & -8.64761419e-02 & 3588.279 & 125891 \\
 & & & & & & \texttt{PNOD} & -8.64761420e-02 & 73.041 & 126391 \\ \hline
 \multirow{3}{*}{D} & \multirow{3}{*}{0} & \multirow{3}{*}{120} & \multirow{3}{*}{12} & \multirow{3}{*}{18} & \multirow{3}{*}{2} & \texttt{Boscia} & -8.03969816e-02 & 381.208 & 5837 \\
 & & & & & & \texttt{Co-BnB} & -8.21119699e-02 & 894.157 & 30855 \\
 & & & & & & \texttt{PNOD} & -8.21119700e-02 & 19.666 & 33535 \\ \hline
 \multirow{3}{*}{D} & \multirow{3}{*}{0} & \multirow{3}{*}{120} & \multirow{3}{*}{12} & \multirow{3}{*}{18} & \multirow{3}{*}{3} & \texttt{Boscia} & -8.54906002e-02 & 772.900 & 11137 \\
 & & & & & & \texttt{Co-BnB} & -8.68811030e-02 & 5424.018 & 176673 \\
 & & & & & & \texttt{PNOD} & -8.68811030e-02 & 113.820 & 181135 \\ \hline
 \multirow{3}{*}{D} & \multirow{3}{*}{0} & \multirow{3}{*}{120} & \multirow{3}{*}{12} & \multirow{3}{*}{18} & \multirow{3}{*}{4} & \texttt{Boscia} & -7.73911467e-02 & 456.244 & 5575 \\
 & & & & & & \texttt{Co-BnB} & -7.90501588e-02 & 10651.424 & 289467 \\
 & & & & & & \texttt{PNOD} & -7.90501590e-02 & 213.481 & 290131 \\ \hline
 \multirow{3}{*}{D} & \multirow{3}{*}{0} & \multirow{3}{*}{120} & \multirow{3}{*}{12} & \multirow{3}{*}{18} & \multirow{3}{*}{5} & \texttt{Boscia} & -8.94308494e-02 & 78.066 & 3011 \\
 & & & & & & \texttt{Co-BnB} & -9.01592955e-02 & 758.279 & 23133 \\
 & & & & & & \texttt{PNOD} & -9.01592950e-02 & 14.862 & 22913 \\ \hline
 \multirow{3}{*}{D} & \multirow{3}{*}{1} & \multirow{3}{*}{50} & \multirow{3}{*}{5} & \multirow{3}{*}{7} & \multirow{3}{*}{1} & \texttt{Boscia} & -4.38946030e-01 &  0.886 & 13 \\
 & & & & & & \texttt{Co-BnB} & -4.38946030e-01 &  0.202 & 21 \\
 & & & & & & \texttt{PNOD} & -4.38946030e-01 &  0.006 & 27 \\ \hline
 \multirow{3}{*}{D} & \multirow{3}{*}{1} & \multirow{3}{*}{50} & \multirow{3}{*}{5} & \multirow{3}{*}{7} & \multirow{3}{*}{2} & \texttt{Boscia} & -4.52556437e-01 &  0.022 & 15 \\
 & & & & & & \texttt{Co-BnB} & -4.52556437e-01 &  1.371 & 47 \\
 & & & & & & \texttt{PNOD} & -4.52556437e-01 &  0.016 & 51 \\ \hline
 \multirow{3}{*}{D} & \multirow{3}{*}{1} & \multirow{3}{*}{50} & \multirow{3}{*}{5} & \multirow{3}{*}{7} & \multirow{3}{*}{3} & \texttt{Boscia} & -4.45269522e-01 &  0.028 & 19 \\
 & & & & & & \texttt{Co-BnB} & -4.45269522e-01 &  0.720 & 35 \\
 & & & & & & \texttt{PNOD} & -4.45269522e-01 &  0.016 & 39 \\ \hline
 \multirow{3}{*}{D} & \multirow{3}{*}{1} & \multirow{3}{*}{50} & \multirow{3}{*}{5} & \multirow{3}{*}{7} & \multirow{3}{*}{4} & \texttt{Boscia} & -4.72192513e-01 &  0.027 & 15 \\
 & & & & & & \texttt{Co-BnB} & -4.72192513e-01 &  1.573 & 67 \\
 & & & & & & \texttt{PNOD} & -4.72192513e-01 &  0.018 & 57 \\ \hline
 \multirow{3}{*}{D} & \multirow{3}{*}{1} & \multirow{3}{*}{50} & \multirow{3}{*}{5} & \multirow{3}{*}{7} & \multirow{3}{*}{5} & \texttt{Boscia} & -4.69818605e-01 &  0.020 & 15 \\
 & & & & & & \texttt{Co-BnB} & -4.69818605e-01 &  0.404 & 45 \\
 & & & & & & \texttt{PNOD} & -4.69818605e-01 &  0.014 & 47 \\ \hline
 \multirow{3}{*}{D} & \multirow{3}{*}{1} & \multirow{3}{*}{60} & \multirow{3}{*}{6} & \multirow{3}{*}{9} & \multirow{3}{*}{1} & \texttt{Boscia} & -4.41930757e-01 &  0.023 & 3 \\
 & & & & & & \texttt{Co-BnB} & -4.54603930e-01 &  0.575 & 55 \\
 & & & & & & \texttt{PNOD} & -4.54603930e-01 &  0.023 & 67 \\ \hline
 \multirow{3}{*}{D} & \multirow{3}{*}{1} & \multirow{3}{*}{60} & \multirow{3}{*}{6} & \multirow{3}{*}{9} & \multirow{3}{*}{2} & \texttt{Boscia} & -4.73992292e-01 &  0.041 & 13 \\
 & & & & & & \texttt{Co-BnB} & -4.74769615e-01 &  1.283 & 113 \\
 & & & & & & \texttt{PNOD} & -4.74769615e-01 &  0.099 & 141 \\ \hline
 \multirow{3}{*}{D} & \multirow{3}{*}{1} & \multirow{3}{*}{60} & \multirow{3}{*}{6} & \multirow{3}{*}{9} & \multirow{3}{*}{3} & \texttt{Boscia} & -4.80148188e-01 &  0.078 & 27 \\
 & & & & & & \texttt{Co-BnB} & -4.80148188e-01 &  6.164 & 275 \\
 & & & & & & \texttt{PNOD} & -4.80148188e-01 &  0.060 & 239 \\ \hline
 \multirow{3}{*}{D} & \multirow{3}{*}{1} & \multirow{3}{*}{60} & \multirow{3}{*}{6} & \multirow{3}{*}{9} & \multirow{3}{*}{4} & \texttt{Boscia} & -5.00814047e-01 &  0.094 & 33 \\
 & & & & & & \texttt{Co-BnB} & -5.01119641e-01 &  6.574 & 393 \\
 & & & & & & \texttt{PNOD} & -5.01119641e-01 &  0.119 & 381 \\ \hline
 \multirow{3}{*}{D} & \multirow{3}{*}{1} & \multirow{3}{*}{60} & \multirow{3}{*}{6} & \multirow{3}{*}{9} & \multirow{3}{*}{5} & \texttt{Boscia} & -4.84853368e-01 &  0.090 & 37 \\
 & & & & & & \texttt{Co-BnB} & -4.85678991e-01 &  7.518 & 385 \\
 & & & & & & \texttt{PNOD} & -4.85678991e-01 &  0.209 & 449 \\ \hline
 \multirow{3}{*}{D} & \multirow{3}{*}{1} & \multirow{3}{*}{80} & \multirow{3}{*}{8} & \multirow{3}{*}{12} & \multirow{3}{*}{1} & \texttt{Boscia} & -5.25732170e-01 &  0.101 & 19 \\
 & & & & & & \texttt{Co-BnB} & -5.26315514e-01 &  5.872 & 311 \\
 & & & & & & \texttt{PNOD} & -5.26390958e-01 &  0.124 & 361 \\ \hline
 \multirow{3}{*}{D} & \multirow{3}{*}{1} & \multirow{3}{*}{80} & \multirow{3}{*}{8} & \multirow{3}{*}{12} & \multirow{3}{*}{2} & \texttt{Boscia} & -4.92815454e-01 &  0.151 & 37 \\
 & & & & & & \texttt{Co-BnB} & -4.93144499e-01 & 12.004 & 717 \\
 & & & & & & \texttt{PNOD} & -4.93144499e-01 &  0.284 & 767 \\ \hline
 \multirow{3}{*}{D} & \multirow{3}{*}{1} & \multirow{3}{*}{80} & \multirow{3}{*}{8} & \multirow{3}{*}{12} & \multirow{3}{*}{3} & \texttt{Boscia} & -4.97196810e-01 &  0.128 & 29 \\
 & & & & & & \texttt{Co-BnB} & -4.99036675e-01 & 10.646 & 689 \\
 & & & & & & \texttt{PNOD} & -4.99036675e-01 &  0.302 & 697 \\ \hline
 \multirow{3}{*}{D} & \multirow{3}{*}{1} & \multirow{3}{*}{80} & \multirow{3}{*}{8} & \multirow{3}{*}{12} & \multirow{3}{*}{4} & \texttt{Boscia} & -5.39280951e-01 &  0.088 & 15 \\
 & & & & & & \texttt{Co-BnB} & -5.39809458e-01 & 17.567 & 887 \\
 & & & & & & \texttt{PNOD} & -5.39809458e-01 &  0.259 & 867 \\ \hline
 \multirow{3}{*}{D} & \multirow{3}{*}{1} & \multirow{3}{*}{80} & \multirow{3}{*}{8} & \multirow{3}{*}{12} & \multirow{3}{*}{5} & \texttt{Boscia} & -5.14781859e-01 &  0.046 & 11 \\
 & & & & & & \texttt{Co-BnB} & -5.17034908e-01 &  9.423 & 467 \\
 & & & & & & \texttt{PNOD} & -5.17034908e-01 &  0.153 & 459 \\ \hline
 \multirow{3}{*}{D} & \multirow{3}{*}{1} & \multirow{3}{*}{100} & \multirow{3}{*}{10} & \multirow{3}{*}{15} & \multirow{3}{*}{1} & \texttt{Boscia} & -5.54143844e-01 &  0.135 & 23 \\
 & & & & & & \texttt{Co-BnB} & -5.55390141e-01 & 43.063 & 2329 \\
 & & & & & & \texttt{PNOD} & -5.55390141e-01 &  1.259 & 2471 \\ \hline
 \multirow{3}{*}{D} & \multirow{3}{*}{1} & \multirow{3}{*}{100} & \multirow{3}{*}{10} & \multirow{3}{*}{15} & \multirow{3}{*}{2} & \texttt{Boscia} & -5.32094833e-01 &  0.387 & 61 \\
 & & & & & & \texttt{Co-BnB} & -5.33834326e-01 & 40.243 & 1943 \\
 & & & & & & \texttt{PNOD} & -5.33834326e-01 &  1.075 & 1943 \\ \hline
 \multirow{3}{*}{D} & \multirow{3}{*}{1} & \multirow{3}{*}{100} & \multirow{3}{*}{10} & \multirow{3}{*}{15} & \multirow{3}{*}{3} & \texttt{Boscia} & -5.40916636e-01 &  0.467 & 71 \\
 & & & & & & \texttt{Co-BnB} & -5.41812190e-01 & 126.122 & 7353 \\
 & & & & & & \texttt{PNOD} & -5.41812190e-01 &  3.483 & 6917 \\ \hline
 \multirow{3}{*}{D} & \multirow{3}{*}{1} & \multirow{3}{*}{100} & \multirow{3}{*}{10} & \multirow{3}{*}{15} & \multirow{3}{*}{4} & \texttt{Boscia} & -5.61916029e-01 &  0.225 & 37 \\
 & & & & & & \texttt{Co-BnB} & -5.62129907e-01 & 99.621 & 4753 \\
 & & & & & & \texttt{PNOD} & -5.62129907e-01 &  2.041 & 4769 \\ \hline
 \multirow{3}{*}{D} & \multirow{3}{*}{1} & \multirow{3}{*}{100} & \multirow{3}{*}{10} & \multirow{3}{*}{15} & \multirow{3}{*}{5} & \texttt{Boscia} & -5.41000284e-01 &  0.243 & 37 \\
 & & & & & & \texttt{Co-BnB} & -5.42561888e-01 & 20.719 & 1035 \\
 & & & & & & \texttt{PNOD} & -5.42561888e-01 &  0.406 & 991 \\ \hline
 \multirow{3}{*}{D} & \multirow{3}{*}{1} & \multirow{3}{*}{120} & \multirow{3}{*}{12} & \multirow{3}{*}{18} & \multirow{3}{*}{1} & \texttt{Boscia} & -5.81582173e-01 &  0.353 & 37 \\
 & & & & & & \texttt{Co-BnB} & -5.82615468e-01 & 794.316 & 31501 \\
 & & & & & & \texttt{PNOD} & -5.82615468e-01 & 16.944 & 29937 \\ \hline
 \multirow{3}{*}{D} & \multirow{3}{*}{1} & \multirow{3}{*}{120} & \multirow{3}{*}{12} & \multirow{3}{*}{18} & \multirow{3}{*}{2} & \texttt{Boscia} & -5.65280433e-01 &  0.490 & 57 \\
 & & & & & & \texttt{Co-BnB} & -5.65280433e-01 & 1301.240 & 70237 \\
 & & & & & & \texttt{PNOD} & -5.65280433e-01 & 34.319 & 71165 \\ \hline
 \multirow{3}{*}{D} & \multirow{3}{*}{1} & \multirow{3}{*}{120} & \multirow{3}{*}{12} & \multirow{3}{*}{18} & \multirow{3}{*}{3} & \texttt{Boscia} & -5.81712313e-01 &  1.242 & 99 \\
 & & & & & & \texttt{Co-BnB} & -5.82068224e-01 & 532.397 & 25741 \\
 & & & & & & \texttt{PNOD} & -5.82068224e-01 & 11.061 & 22717 \\ \hline
 \multirow{3}{*}{D} & \multirow{3}{*}{1} & \multirow{3}{*}{120} & \multirow{3}{*}{12} & \multirow{3}{*}{18} & \multirow{3}{*}{4} & \texttt{Boscia} & -5.85436610e-01 &  0.229 & 25 \\
 & & & & & & \texttt{Co-BnB} & -5.86172636e-01 & 233.831 & 12003 \\
 & & & & & & \texttt{PNOD} & -5.86172636e-01 &  5.632 & 12347 \\ \hline
 \multirow{3}{*}{D} & \multirow{3}{*}{1} & \multirow{3}{*}{120} & \multirow{3}{*}{12} & \multirow{3}{*}{18} & \multirow{3}{*}{5} & \texttt{Boscia} & -5.75351838e-01 &  0.422 & 41 \\
 & & & & & & \texttt{Co-BnB} & -5.77930932e-01 & 43.764 & 2351 \\
 & & & & & & \texttt{PNOD} & -5.77930932e-01 &  1.020 & 2419 \\ \hline
\end{longtable}

\bibliographystyle{alpha}
\bibliography{refs}

\end{document}